\documentclass[prd,twocolumn,superscriptaddress,floatfix,nopacs,preprintnumbers,nofootinbib]{revtex4-2}
\usepackage[utf8]{inputenc}

\usepackage[caption = false]{subfig}
\usepackage[normalem]{ulem}
\usepackage{xcolor}

\usepackage{mathrsfs}
\usepackage{amssymb,bm}
\usepackage{amsmath}
\usepackage{mathtools}
\usepackage{physics}
\usepackage{slashed}

\newcommand{\aem}{\alpha_\mathrm{em}}
\newcommand{\gev}{\mathrm{GeV}}

\newcommand{\bt}{\mathbf{b}}
\newcommand{\rt}{\mathbf{r}}

\usepackage[breaklinks,colorlinks,citecolor=citcolor,urlcolor=blue,linkcolor=lcolor]{hyperref}

\definecolor{lcolor}{rgb}{0.5,0,0}
\definecolor{citcolor}{rgb}{0,0.3,0.0}

\newcommand{\as}{\alpha_\mathrm{s}}

\begin{document}

\author{Nestor Armesto}
\email{nestor.armesto@usc.es}
\affiliation{Instituto Galego de Física de Altas Enerxías IGFAE,
Universidade de Santiago de Compostela, 15782 Santiago de Compostela, Galicia-Spain}
\author{Tuomas Lappi}
\email{tuomas.v.v.lappi@jyu.fi}
\author{Heikki Mäntysaari}
\email{heikki.mantysaari@jyu.fi}
\author{Hannu Paukkunen}
\email{hannu.paukkunen@jyu.fi}
\affiliation{
Department of Physics, University of Jyväskylä,  P.O. Box 35, 40014 University of Jyväskylä, Finland
}
\affiliation{
Helsinki Institute of Physics, P.O. Box 64, 00014 University of Helsinki, Finland
}
\author{Mirja Tevio}
\email{mirja.h.tevio@jyu.fi}
\affiliation{
Department of Physics, University of Jyväskylä,  P.O. Box 35, 40014 University of Jyväskylä, Finland
}
\affiliation{
Helsinki Institute of Physics, P.O. Box 64, 00014 University of Helsinki, Finland
}

\title{Signatures of gluon saturation from structure-function measurements}

\begin{abstract}

We study experimentally observable signals for  nonlinear QCD dynamics in deep inelastic scattering (DIS) at small
Bjorken variable $x$ and moderate virtuality $Q^2$, by quantifying differences between the linear Dokshitzer-Gribov-Lipatov-Altarelli-Parisi (DGLAP) evolution and nonlinear evolution with the Balitsky-Kovchegov (BK) equation.
To remove the effect of the parametrization freedom in the initial conditions of both equations, we first match the predictions for the DIS structure functions $F_2$ and $F_{\rm L}$ from both frameworks in a region in $x,Q^2$ where both frameworks should provide an accurate description of the relevant physics. The differences in the dynamics are then quantified by the deviations when one moves away from this matching region.
For free protons we find that the differences in $F_2$ remain at a few-percent level, while in $F_{\rm L}$ the deviations are larger, up to $10\,\%$ at the EIC and $40\,\%$ at the LHeC kinematics. With a heavy nucleus the differences are up to $10\,\%$ in $F_2$, and can reach $20\,\%$ and $60\,\%$ in $F_{\rm L}$ for the EIC and the LHeC, respectively.

\end{abstract}

\maketitle

\section{Introduction}

The high-energy regime of Quantum Chromodynamics (QCD) is characterised by high partonic densities where non-linear phenomena play a dominant role~\cite{Kovchegov:2012mbw}. In this regime, the predictions of usual linear evolution equations for parton densities like the Dokshitzer-Gribov-Lipatov-Altarelli-Parisi (DGLAP) equation \cite{Gribov:1972ri,Gribov:1972rt,Altarelli:1977zs,Dokshitzer:1977sg} which provides an accurate description of the QCD dynamics at moderate to large values of the momentum fractions $x$ of the probed parton and virtualities $Q^2\gg\Lambda_{\rm QCD}^2$, or the Balitsky-Fadin-Kuraev-Lipatov (BFKL) equation \cite{Kuraev:1977fs,Balitsky:1978ic} which accurately describes the partonic dynamics at small $x$, are expected to become less reliable. High-density effects should become relevant and unitarity corrections or, alternatively, gluon recombination processes, tame the growth of parton densities towards small $x$ and make them saturate -- thus non-linear dynamics is also referred to as saturation. Non-linear modifications  to DGLAP evolution equations were first proposed in Refs.~\cite{Gribov:1983ivg,Mueller:1985wy}. Later on, a formulation suitable for QCD at such high parton densities and energies (or small $x$) was developed as a weak coupling but non-perturbative effective field theory, the Color Glass Condensate (CGC), see Refs.~\cite{Kovchegov:2012mbw,Gelis:2010nm}. Such approach resums all higher-twist corrections that are neglected in linear DGLAP evolution. In this framework, the BFKL equation for evolution in $x$ is generalized to the Balitsky-Kovchegov one (BK)~\cite{Balitsky:1995ub,Kovchegov:1999yj,Kovchegov:1999ua}. The BK equation includes a nonlinear term needed to preserve the unitarity of the (dipole) scattering amplitude.

Finding such a novel regime of QCD would be of uttermost importance for  understanding QCD, for developing more realistic initial conditions for simulations of heavy ion collisions at high energy and for computing the main backgrounds for signals of physics beyond the Standard Model in hadronic collisions, especially at the high energies of the Future Circular Collider~\cite{FCC:2018byv,FCC:2018vvp}.
Linear, fixed-order DGLAP evolution equations are known to provide a very good description of the structure of the proton at the available values of $x$ and $Q^2$, see e.g. Ref.~\cite{Gao:2017yyd}. However, there have also been claims that the  HERA data~\cite{H1:2009pze,H1:2015ubc} at small $x$ and moderate $Q^2$ necessitates the inclusion of small-$x$ resummation~\cite{Ball:2017otu,xFitterDevelopersTeam:2018hym} or non-linear evolution~\cite{Albacete:2012rx}.\footnote{The impact of the former on Higgs production at hadronic colliders on Higgs production has been examined in~\cite{Bonvini:2018iwt}.} Note also that while DGLAP-based fits provide a good description of available data constraining parton densities inside nuclei (nPDFs), see e.g.~\cite{Kovarik:2015cma,Khanpour:2016pph,Eskola:2016oht,Helenius:2021tof,AbdulKhalek:2020yuc,Eskola:2021nhw}, non-linear effects are density effects that should be enhanced by the nuclear size. Therefore, searches for saturation with heavier nuclei are crucial not only because the expected effects are stronger, and thereby more easily detectable, but also to eventually check the theoretical explanation of such non-linear dynamics, if found. For the moment, however, even the D-meson production at the LHC that can probe the nuclei down to $x \sim 10^{-6}$ at low $Q^2$ shows no visible deviation from the DGLAP approach~\cite{Eskola:2019bgf,Eskola:2021nhw}.    

Searches of saturation are ongoing in hadronic colliders~\cite{Morreale:2021pnn} -- the Relativistic Heavy Ion Collider (RHIC) at BNL and the Large Hadron Collider (LHC) at CERN, see e.g.~\cite{Citron:2018lsq} for future prospects at the LHC. However, with a cleaner environment and direct experimental access to the kinematical variables $x,Q^2$, deep inelastic scattering (DIS) in lepton-proton/nucleus  collisions offers the ideal environment to study the partonic structure of protons and heavier nuclei. The currently projected and proposed accelerators, the Electron Ion Collider~\cite{Accardi:2012qut,AbdulKhalek:2021gbh} (EIC) at BNL, and the Large Hadron-electron Collider~\cite{LHeCStudyGroup:2012zhm,LHeC:2020van} (LHeC) and the Future Circular Collider in electron-hadron mode~\cite{FCC:2018byv,FCC:2018vvp} (FCC-he) at CERN, will provide lepton-hadron collisions in the centre-of-mass range from a few tens of GeV to a few TeV per nucleon. Together with the use of new detector techniques, they promise to revolutionise our understanding of hadron structure and partonic dynamics.

In DIS, several observables have been proposed as being sensitive to non-linear dynamics: particle correlations~\cite{Marquet:2009ca,LHeCStudyGroup:2012zhm}, diffraction~\cite{Accardi:2012qut,Mantysaari:2017slo}, and deviations from linear evolution in inclusive observables, in particular  the total cross section.
Concerning the latter, it is expected that the EIC and the LHeC will be sensitive to saturation, i.e. to deviations from DGLAP due to higher-twist effects. However, conclusively distinguishing these two pictures using experimental data can, in practice, be difficult. This is due to the fact that both non-linear and DGLAP-based calculations require non-perturbative inputs that are obtained by fitting experimental data. When the initial conditions of the evolution are fit to the same data, also the predictions will not deviate dramatically. Thus  different theoretical frameworks might be equally capable of describing experimental data \emph{a posteriori}.
Additionally, the DGLAP and BK equations predict the evolution of cross sections in different directions: increasing $Q^2$ and decreasing $x$, respectively. Thus one cannot solve both starting from the same initial condition: the perturbative QCD prediction of one equation is the nonperturbative initial condition of the other.  Therefore, it is important to be careful to distinguish between genuine effects of the different evolution dynamics on one hand, and the ability of the initial condition parametrizations to adjust to data on the other hand.

Several works have investigated the possibility to disentangle DGLAP linear dynamics from saturation in $ep$ collisions at the LHeC~\cite{Rojo:2009ut,LHeCStudyGroup:2012zhm,LHeC:2020van} and $eA$ collisions at the EIC~\cite{Marquet:2017bga}. The basic technique employed in these studies is the inclusion of pseudodata from the projected experiments in a DGLAP-based global fit. Then, some measure of the fit quality is computed. Generically, a worsening is observed when the introduced pseudodata are generated using a model that contains non-linear dynamics in comparison to the case when the corresponding pseudodata are produced using linear DGLAP dynamics. 
Focusing on the difference between the generated pseudodata and existing DGLAP-evolved parton distributions, as in Ref.~\cite{Marquet:2017bga}, addresses potential tensions between predictions from nonlinear dynamics and \emph{existing} PDF parametrizations. The latter are determined not only by the QCD dynamics, but  by existing data and to some extent by parametrization choices. Thus the question addressed by such a comparison has many more facets than the simple one we pose here, namely: how different is the actual evolution dynamics  once one allows both fits to adjust themselves to a common set of data?

The purpose of this paper is to evaluate the difference between DGLAP and non-linear evolution in the kinematic regions corresponding to the EIC and the LHeC, both for $ep$ and $eA$ collisions, with a slightly different emphasis than previous studies. Our procedure is the following.  We first produce results for the $F_2$ and $F_{\rm L}$ structure functions  using non-linear evolution equations. Then, employing reweighting techniques, a DGLAP evolved set of parton densities is produced that results in structure functions coinciding with the non-linear evolution ones in a line of the $x,Q^2$ plane, \emph{as precisely as possible}. Technically this is achieved by a Bayesian reweighting process. Here one takes an existing collection  of ``replica'' PDF sets that all satisfy the DGLAP equation, and calculates for each of them a weight defined by how close it is to the given $F_2$ and $F_{\rm L}$ structure functions.  The reweighted structure functions and PDFs are then defined as a weighted sum over all replicas. We will explain this procedure in more detail below in  Sec.~\ref{sec:reweighting}.
Now the difference between the non-linear results and the DGLAP ones for $F_2$ and $F_{\rm L}$ when moving away from this line  provides a quantification  of the signatures of  saturation -- of the difference between linear and non-linear dynamics. Although we are interested in the experimentally accessible kinematical regimes, we do not use estimated experimental errors in this procedure.

Our approach can be compared to the one used in the previous studies
\cite{Rojo:2009ut,LHeCStudyGroup:2012zhm,LHeC:2020van,Marquet:2017bga}, where a DGLAP fit to pseudodata from BK-evolution in the whole kinematical region was performed. There one fits to the pseudodata in a wide region in the kinematical plane, and the weight given to specific kinematical regions is determined by the projected errors. Our goal, in contrast, is to force the two approaches to agree quite precisely on just a line in the kinematical plane, preferably in a region where both theoretical frameworks should be valid. The  deviations as one moves away from this line provide a clearer picture of the genuine differences in the evolution dynamics, independently of experimental errors, kinematical ranges or PDF parametrizations.

The structure of the manuscript is as follows. In Section~\ref{sec:collinear} we review the method to compute the structure functions in collinear factorisation that will be employed to produce the DGLAP results. Section~\ref{sec:saturation} addresses the same point in the framework of non-linear, saturation dynamics. The PDF reweighting technique to produce a DGLAP fit coincident with the saturation results in a line of the $x,Q^2$ plane is presented in Section~\ref{sec:reweighting}. Results for proton and nucleus are shown and discussed in Section~\ref{sec:results}. Finally, we present our conclusions in Section~\ref{sec:conclusions}.

\section{Theoretical Framework}
\label{sec:thsetup}

\subsection{Structure functions in collinear factorization}
\label{sec:collinear}

To compute the structure functions in collinear factorization we use APFEL~\cite{Bertone:2013vaa} which accesses the PDFs through the LHAPDF library~\cite{Buckley:2014ana}. For the proton we use the next-to-leading order (NLO) NNPDF3.1 set \texttt{ NNPDF31\_nlo\_as\_0118\_1000}~\cite{NNPDF:2017mvq} which has $N_{\rm rep}=1000$ replicas describing the PDF uncertainty, and for gold (our representative of a heavier nucleus) we use the NLO nNNPDF2.0 \texttt{nNNPDF20\_nlo\_as\_0118\_Au197}~\cite{AbdulKhalek:2020yuc} set which provides 1000 replicas as well\footnote{We note that a preprint of the updated nNNPDF3.0 set~\cite{Khalek:2022zqe} became available when this work was being finalized. }. The central value for any PDF-dependent quantity $\mathcal{O}$ is obtained as an average over the individual predictions $\mathcal{O}[f_k]$ for each individual replica,
\begin{equation}
    \mathcal{O} = \frac{1}{N_{\rm rep}}\sum_{k=1}^{N_{\rm rep}}\mathcal{O}[f_k]
\end{equation}
and the uncertainty $\delta\mathcal{O}$ is computed as the variance, 
\begin{equation}
    \delta\mathcal{O} = \sqrt{\frac{1}{N_{\rm rep}}\sum_{k=1}^{N_{\rm rep}}\left(\mathcal{O}[f_k]-\mathcal{O} \right)^2}.
\end{equation}

We calculate the neutral-current DIS structure functions at NLO accuracy in the FONLL-B general-mass scheme~\cite{Forte:2010ta}. This scheme matches with the one used in the actual global analyses. 
With the NNPDF3.1 proton PDF set we ensure the consistent treatment of fitted charm quarks by setting the appropriate options in APFEL~\cite{Ball:2015tna}, in particular enabling intrinsic charm by \texttt{APFEL::EnableIntrinsicCharm(true)}.  
All the parameters (masses of heavy quarks, values of strong coupling $\alpha_{\rm s}$, etc...) are set to be the same as in the corresponding PDF fits. 

We note that the NNPDF collaboration has also prepared PDF sets including small-$x$ resummation effects at leading and next-to-leading logarithmic accuracy~\cite{Ball:2017otu}. In particular the (linear) small-$x$ resummation was found to be necessary to describe the HERA structure function data in the small-$x$, moderate $Q^2$ region at next-to-next-to-leading order fits. At NLO accuracy, on the other hand, this resummation does not have a numerically significant effect on structure functions at HERA kinematics and is not considered in this work.

\subsection{Structure functions with gluon saturation}
\label{sec:saturation}

In order to calculate proton and nuclear structure functions taking into account non-linear saturation effect we use the Color Glass Condensate (CGC) framework~\cite{Kovchegov:2012mbw,Iancu:2003xm,Gelis:2010nm}. Here it is convenient to describe DIS  processes in the dipole picture in a frame where the target is at rest and the photon plus light-cone momentum is large, and write the leading order total photon-nucleus cross section as
\begin{equation}
\label{eq:cgc-gammap}
\sigma^{\gamma^* A}_{T,L} = 2\sum_f \int \dd[2]{\bt} \dd[2]{\rt} \dd{z} \left|\Psi^{\gamma^* \to q\bar q}(\rt,z,Q^2)\right|^2 N(\bt,\rt,x).
\end{equation}
Here the photon light front wave function $\Psi^{\gamma^* \to q\bar q}(\rt,z,Q^2)$ describes the photon splitting to a quark (with flavor $f$) dipole with transverse size $\rt$ and the quark carrying a fraction $z$ of the photon plus momentum~\cite{Kovchegov:2012mbw}. All information about the target is encoded in the dipole-target scattering amplitude $N(\rt,\bt,x)$, which describes the eikonal propagation of the quark-antiquark pair through the target color field. The subscripts $T$ and $L$ refer to the photon polarization states.  The structure functions can then be written as
\begin{align}
F_2(x,Q^2) &= \frac{Q^2}{4\pi \aem}\left(\sigma^{\gamma^* A}_{T} + \sigma^{\gamma^* A}_{L} \right), \\
F_L(x,Q^2) &= \frac{Q^2}{4\pi \aem}\sigma^{\gamma^* A}_{L}. 
\end{align}

The energy (or Bjorken-$x$) evolution of the dipole-target scattering amplitude $N$ (and thus the $x$ dependence of the structure functions) can be obtained by solving the perturbative Balitsky-Kovchegov (BK) evolution equation~\cite{Kovchegov:1999yj,Balitsky:1995ub} which resums contributions $\sim \as \ln 1/x$ to all orders. In contrast to the collinear factorization based approach described above, the $Q^2$ dependence  is not  calculated perturbatively -- at least at leading order. 
Instead, it is effectively determined, along with other parameters, by the non-perturbative initial condition for the BK evolution. These initial conditions are obtained by fitting the HERA structure function data~\cite{H1:2009pze,H1:2015ubc}. In this work we use the ``MV$^e$'' leading order fit with running coupling corrections~\cite{Balitsky:2006wa} from Ref.~\cite{Lappi:2013zma} (see also a similar fit in  Ref.~\cite{Albacete:2010sy}).

One now needs to  generalize the dipole-proton scattering amplitude to the dipole-nucleus case in order to calculate  nuclear structure functions. We again do this using the procedure of Ref.~\cite{Lappi:2013zma}, where the squared nuclear saturation scale is taken to be proportional to the transverse density $T_A$ as $Q_{s,A}^2 \sim T_A(\bt)$. This scaling is used to generate the initial condition for the BK evolution separately at every impact parameter $\bt$, and different impact parameters are then evolved independently. 

In principle it would be possible to include impact parameter dependence also in the BK evolution, but that would require us to include an additional phenomenological description of confinement effects, see e.g. Refs.~\cite{Berger:2012wx,Berger:2011ew,Berger:2010sh,Mantysaari:2018zdd}. As the structure functions are only sensitive to the integral over the impact parameter, we expect additional effects from the geometry evolution to only have a small effect on our results.  In the region where the saturation scale of the nucleus would fall below that of the proton we use, again following Ref.~\cite{Lappi:2013zma}, a dipole-proton scattering amplitude scaled such that the total dipole-nucleus cross section would be simply $A$ times the dipole-proton cross section. By construction the nuclear modifications then vanish in the dilute region at the  edge of the nucleus. This and similar setups have been successfully used to describe various LHC and RHIC measurements~\cite{Mantysaari:2019nnt,Ducloue:2016pqr,Ducloue:2015gfa,Lappi:2012nh,Ma:2017rsu,Fujii:2013gxa,Tribedy:2010ab}.

Equation \eqref{eq:cgc-gammap} corresponds to the leading order contribution (with $\as \ln 1/x$ contributions resummed to all orders). Currently there is a rapid progress in the field towards next-to-leading order (NLO) accuracy, and necessary ingredients for NLO phenomenology are becoming available. These include, for example, the virtual photon wave function~\cite{Beuf:2021srj,Beuf:2020dxl,Beuf:2021qqa,Hanninen:2017ddy,Beuf:2017bpd,Beuf:2016wdz} and the BK evolution equation~\cite{Lappi:2020srm,Lappi:2016fmu,Lappi:2015fma,Balitsky:2008zza} at NLO. In the first phenomenological structure function calculations at NLO accuracy~\cite{Beuf:2020dxl}, very small differences were found between running coupling LO and NLO calculations in the HERA kinematics when the non-perturbative initial condition for the BK evolution is fitted to the data.
This means that in leading order setups such as the one used in this work the fit parameters can effectively capture most of the higher order corrections. Consequently we expect our leading order framework to be sufficient for the purposes of this work, where the aim is to study differences between the DGLAP and BK dynamics.

\subsection{Matching}
\label{sec:reweighting}

As a technical method to match the structure functions in collinear factorization to those calculated with the BK evolution as discussed in Sec.~\ref{sec:saturation}, we use the Bayesian reweighting method \cite{Ball:2010gb,Ball:2011gg}. The matching is achieved by reweighting the PDFs  to reproduce structure functions calculated from the BK equation in a line in the $x,Q^2$-plane. The preferred region for this matching is the region where both theoretical frameworks should be valid. Thus one wants to be sufficiently above the saturation scale $Q_s^2$ to be in a linear regime described by DGLAP, but not so high that the large logarithms of $Q^2$, not resummed by BK evolution, dominate. An additional technical requirement is that the matching points should be above the minimum scale $Q_0^2$ for which the PDF sets that we use are available. Based on these considerations we have chosen to match the BK and collinear factorization calculations in the  region $10Q_s^2(x) < Q^2 < 11 Q_s^2(x)$, where $Q_s^2(x)$ is the saturation scale of the proton as defined in Ref.~\cite{Lappi:2013zma}. This leads to the matching interval $x <  5.6\times10^{-3}$ for the proton and $x < 1\times10^{-2}$ for the gold nucleus. We perform the matching above a lower limit  $x>1.03\times10^{-5}$ that is chosen to be in the LHeC kinematical regime.  In our matching region the saturation effects are expected to be weak, but $Q^2$ is still not too large. Consequently the DGLAP- and BK-evolution based frameworks are expected to deviate from each other less dramatically than if the matching was performed in some more extreme region of phase space.
Our results naturally depend on the choice of the matching scale; the results  obtained with a matching at a  higher value of $Q^2$ are discussed in Appendix~\ref{app:highQmatch}.

Matching the DGLAP and BK predictions using reweighting amounts to forming an appropriate linear combination of the available PDF replicas such that the predictions given by this linear combination reproduce the structure functions from the BK framework. In practice, for each PDF replica $f_k$ we define a figure-of-merit $\chi_k^2$ by 
\begin{equation}
\label{eq:chi2}
    \chi^2 = \sum_{i,j=1}^{N_{\rm data}} (y_i-y_i[f_k])\sigma^{-1}_{ij}(y_j-y_j[f_k]),
\end{equation}
where $y_i$ denote the $N_{\rm data}$ matching values from BK and $y_i[f_k]$ are the  corresponding values calculated by using a given PDF replica. The order of magnitude of $\chi^2$ can be controlled by the entries of the covariance matrix $\sigma_{ij}$ which is taken to be of the diagonal form
\begin{equation}
\label{eq:sigma_ij}
    \sigma_{ij}=(\delta_{\rm BK}y_i)^2\delta_{ij},
\end{equation} 
where $\delta_{\rm BK}$ is a constant. In a sense, we are thus assigning a constant relative uncertainty on the predictions (``pseudodata'') generated from the BK setup.

For each replica we assign the so-called Giele-Keller weight \cite{Giele:1998gw},
\begin{equation}
\label{eq:omega}
    \omega_k = \frac{e^{-\frac{1}{2}\chi^2_k}}{\frac{1}{N_{\rm rep}}\sum_{k=1}^{N_{\rm rep}}e^{-\frac{1}{2}\chi^2_k}},
\end{equation}
which always sum up to unity, 
\begin{equation}
\frac{1}{N_{\rm rep}} \sum_{k=1}^{N_{\rm rep}}\omega_k = 1 \,.
\end{equation}
The value for any quantity $\mathcal{O}$ using the reweighted PDFs is then obtained as a weighted average, 
\begin{equation}
    \label{eq:reweighted_obs}
    \mathcal{O}^{\rm Rew} = \frac{1}{N_{\rm rep}}\sum_{k=1}^{N_{\rm rep}}\omega_k\mathcal{O}[f_k]  
\end{equation}
with an uncertainty estimate
\begin{equation}
    \label{eq:reweighted_obs_err}
    \delta\mathcal{O}^{\rm Rew} = \sqrt{\frac{1}{N_{\rm rep}}\sum_{k=1}^{N_{\rm rep}}\omega_k\left(\mathcal{O}[f_k]-\mathcal{O}^{\rm Rew} \right)^2}.
\end{equation}

We note that in the case of reweighting to real experimental data, the appropriate weights to be used in conjunction with the NNPDF replicas include also a term $(\chi^2_k)^{(N_{\rm data}-1)/2}$ \cite{Ball:2010gb,Paukkunen:2014zia}. While the Giele-Keller weights favor replicas with $\chi_k^2/N_{\rm data}\approx 0$, the other weights favor replicas with $\chi^2_k/N_{ \rm data} \approx 1$, as discussed in Ref. \cite{Paukkunen:2014zia}. Here we are matching PDFs to smooth values resulting from a theory calculation, without point-by-point statistical fluctuations, not reweighting to experimental data. Thus the preferred values of $\chi_k^2/N_{\rm data}$ should be close to 0 and the Giele-Keller weights are better suited to our purpose.

\begin{figure*}[tbh!]
\label{fig:F2FLproton}
    \centering
    \subfloat[$F_2$\label{subfig:F2protonF2weight}]{%
        \includegraphics[width=\columnwidth]{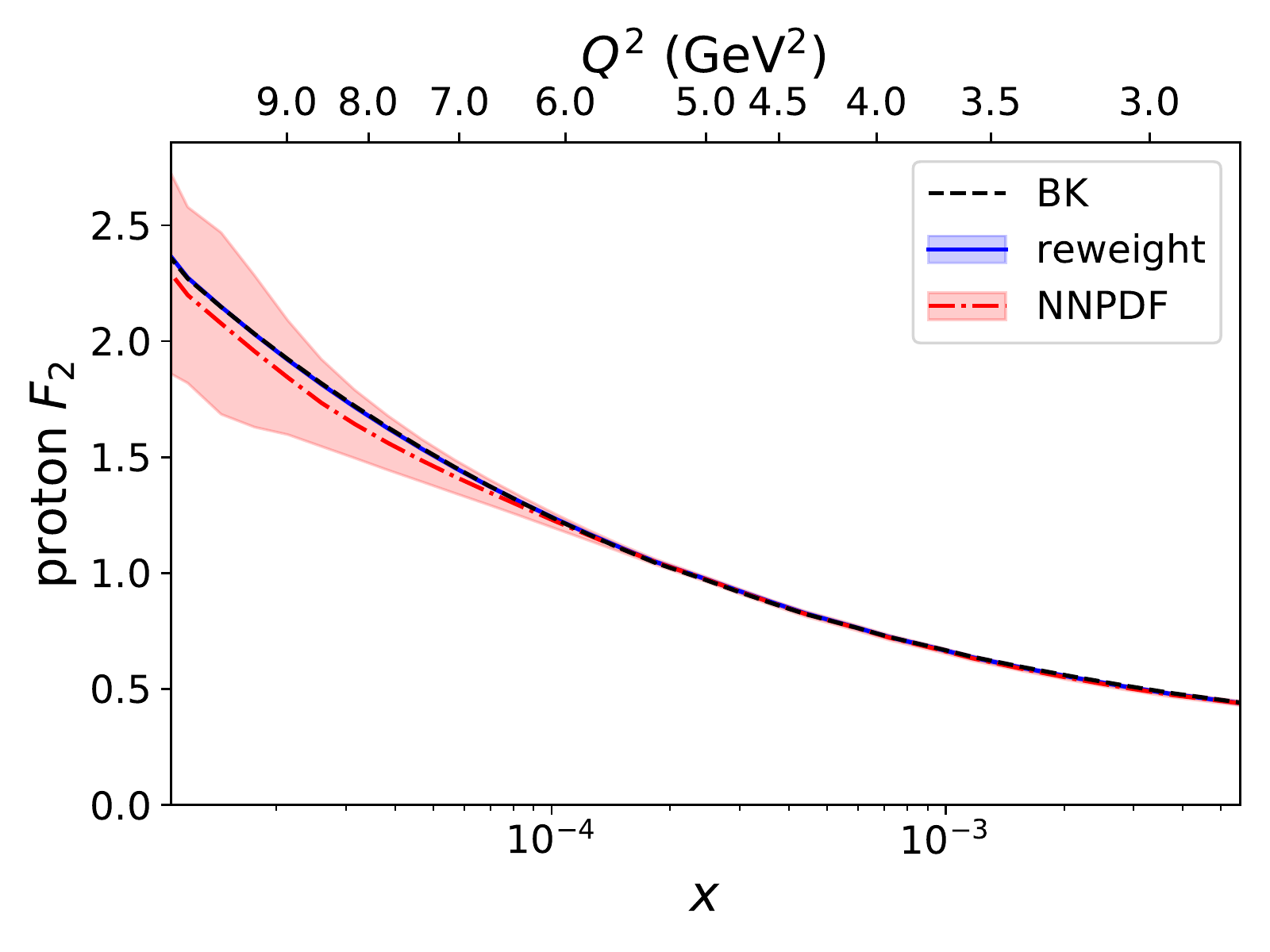}
    }
    \hfill
    \subfloat[$F_{\rm L}$\label{subfig:FLprotonFLweight}]{%
        \includegraphics[width=\columnwidth]{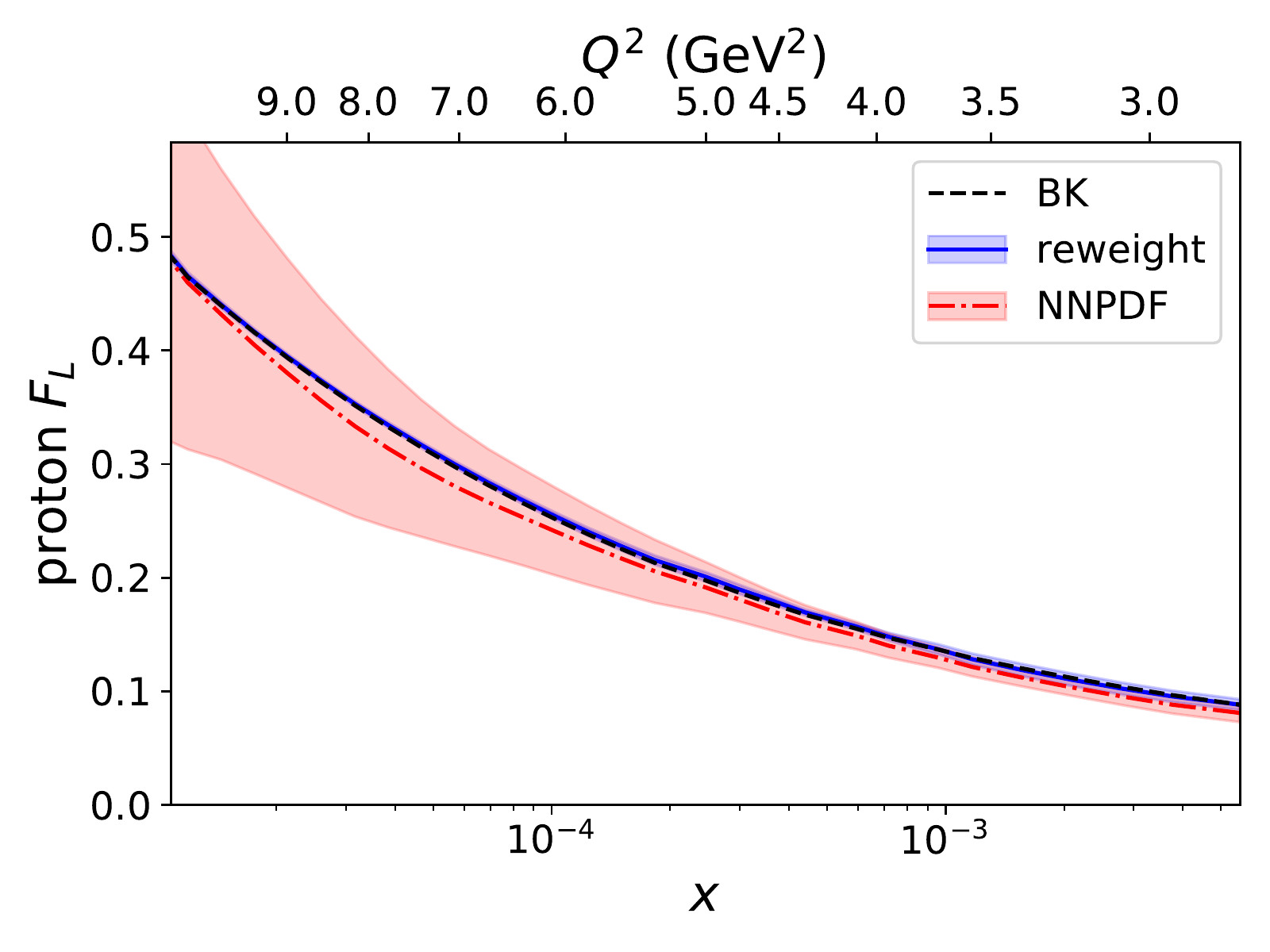}
    } 
    \caption{The $F_2$ (a) and $F_{\rm L}$ (b) structure functions for proton as a function of $x$ at $Q^2=10Q_s^2(x)$. The black dashed curve shows the BK predictions, the red dashed-dotted curve with the red error band the original NNPDF3.1 PDF predictions, and the blue solid curve with a light-blue errorband (too narrow to be visible) the PDF predictions after the matching. 
        }
\end{figure*}

\begin{figure*}[tbh!]
\centering
    \subfloat[$F_2$\label{subfig:HmapF2protonF2weight}]{%
        \includegraphics[width=\columnwidth]{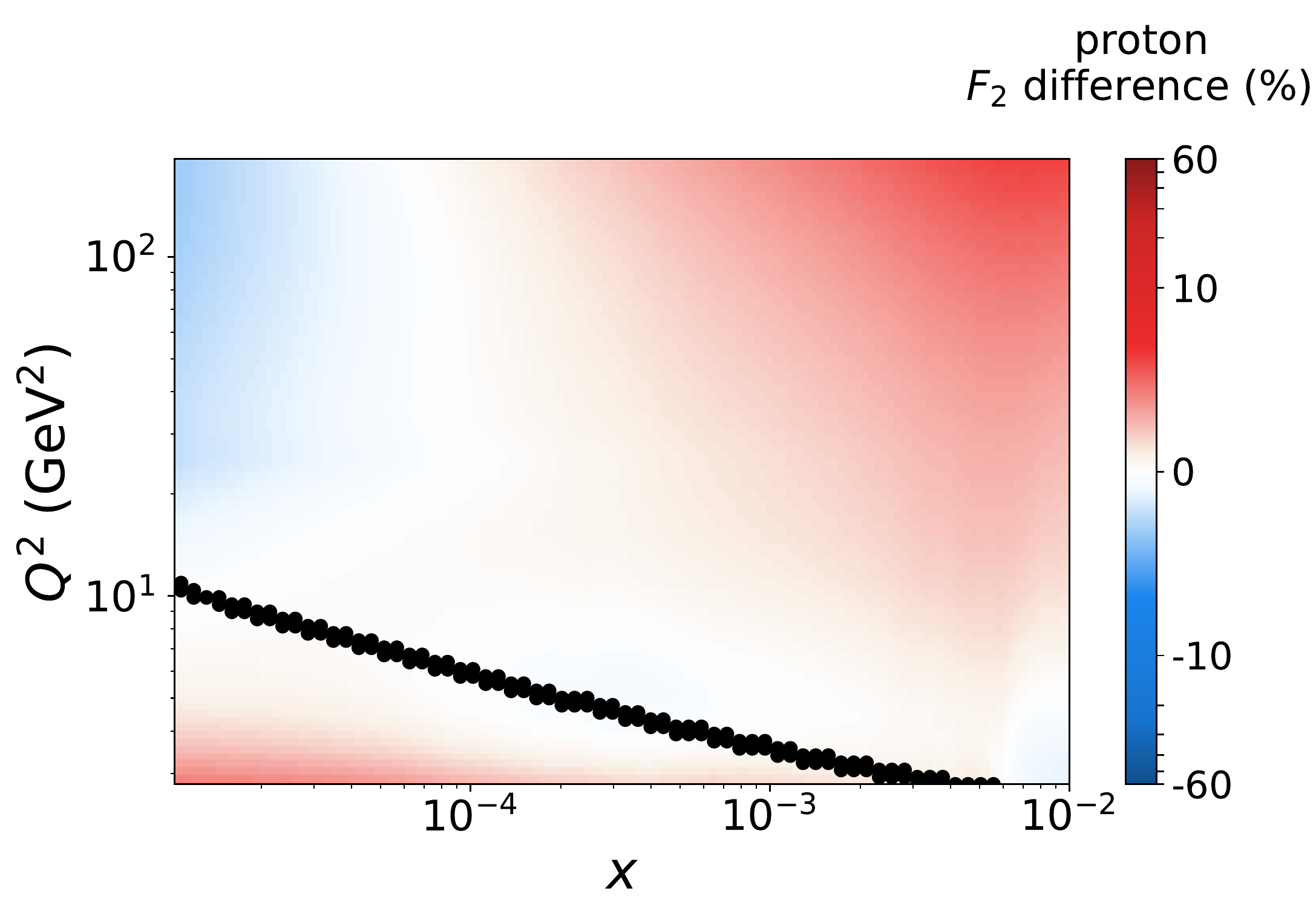}
    }
    \hfill
    \subfloat[$F_{\rm L}$\label{subfig:HmapFLprotonFLweight}]{%
        \includegraphics[width=\columnwidth]{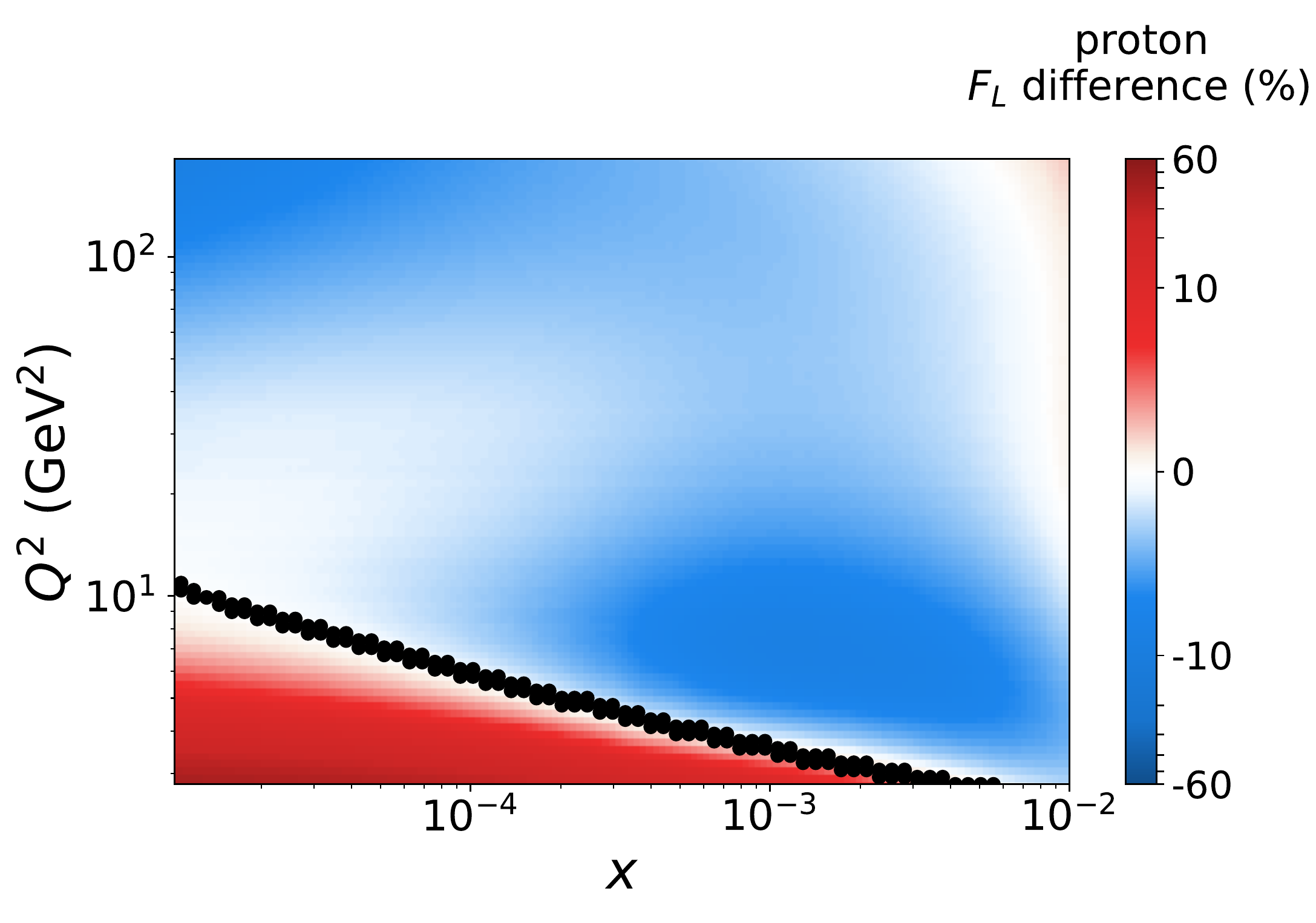}
    }
\caption{Relative difference $(F_{2,L}^{\rm BK}-F_{2,L}^{\rm Rew})/F_{2,L}^{\rm BK}$ between the BK structure functions and the matched $F_2$ (a) and $F_{\rm L}$ (b) for proton as a function of $x$ and $Q^2$. The color scale/axis goes in a linear scale from $-10\, \%$ to $10\, \%$ and in a logarithmic scale outside that range. The black dots indicate the matching points.} 
\label{fig:HmapF2FLproton}
\end{figure*}

\begin{figure*}[tbh!]
\centering
\subfloat[$F_2$\label{subfig:1DF2protonF2weight}]{%
    \includegraphics[width=\columnwidth]{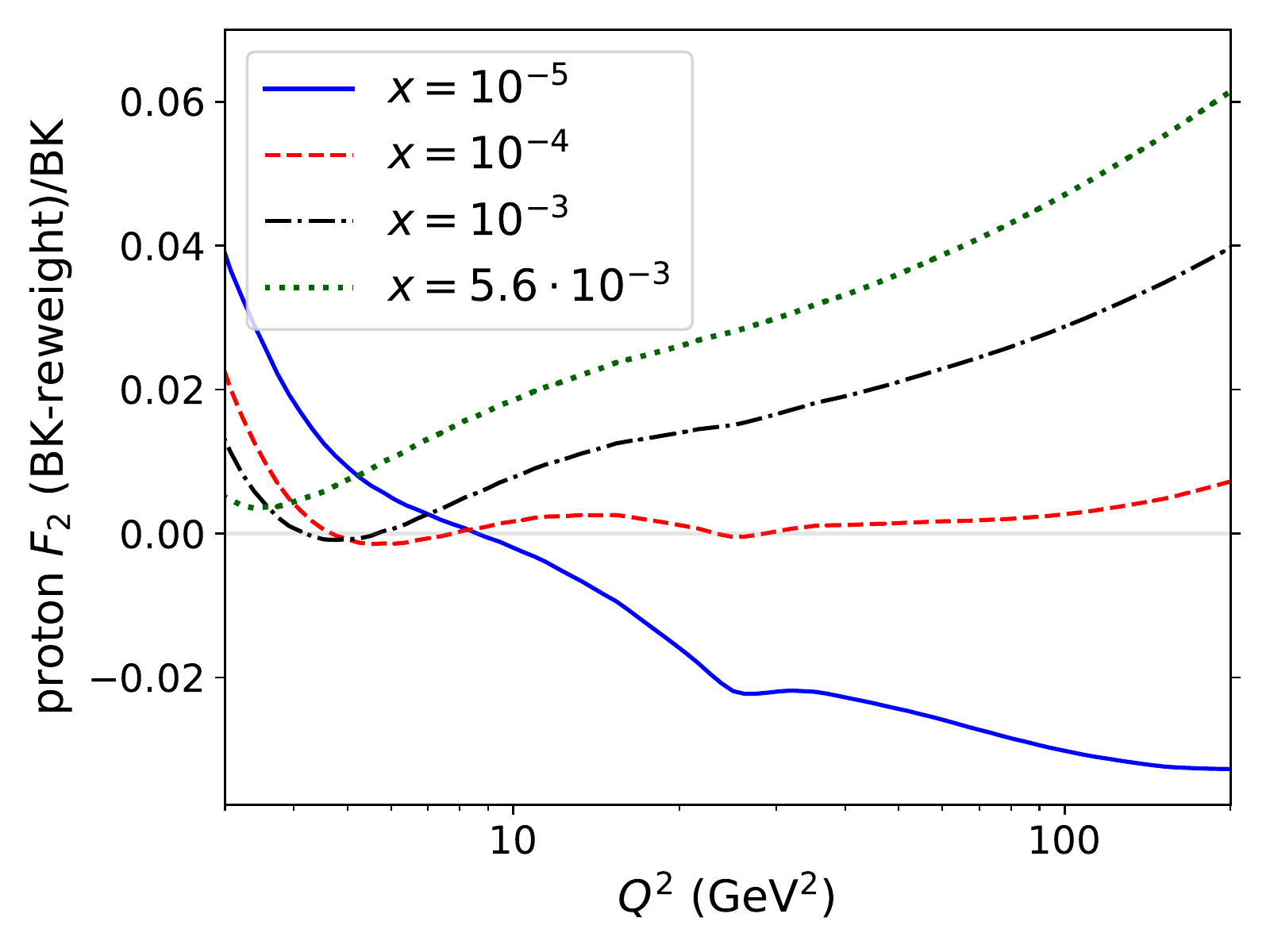}
}
\hfill
\subfloat[$F_{\rm L}$\label{subfig:1DFLprotonFLweight}]{%
    \includegraphics[width=\columnwidth]{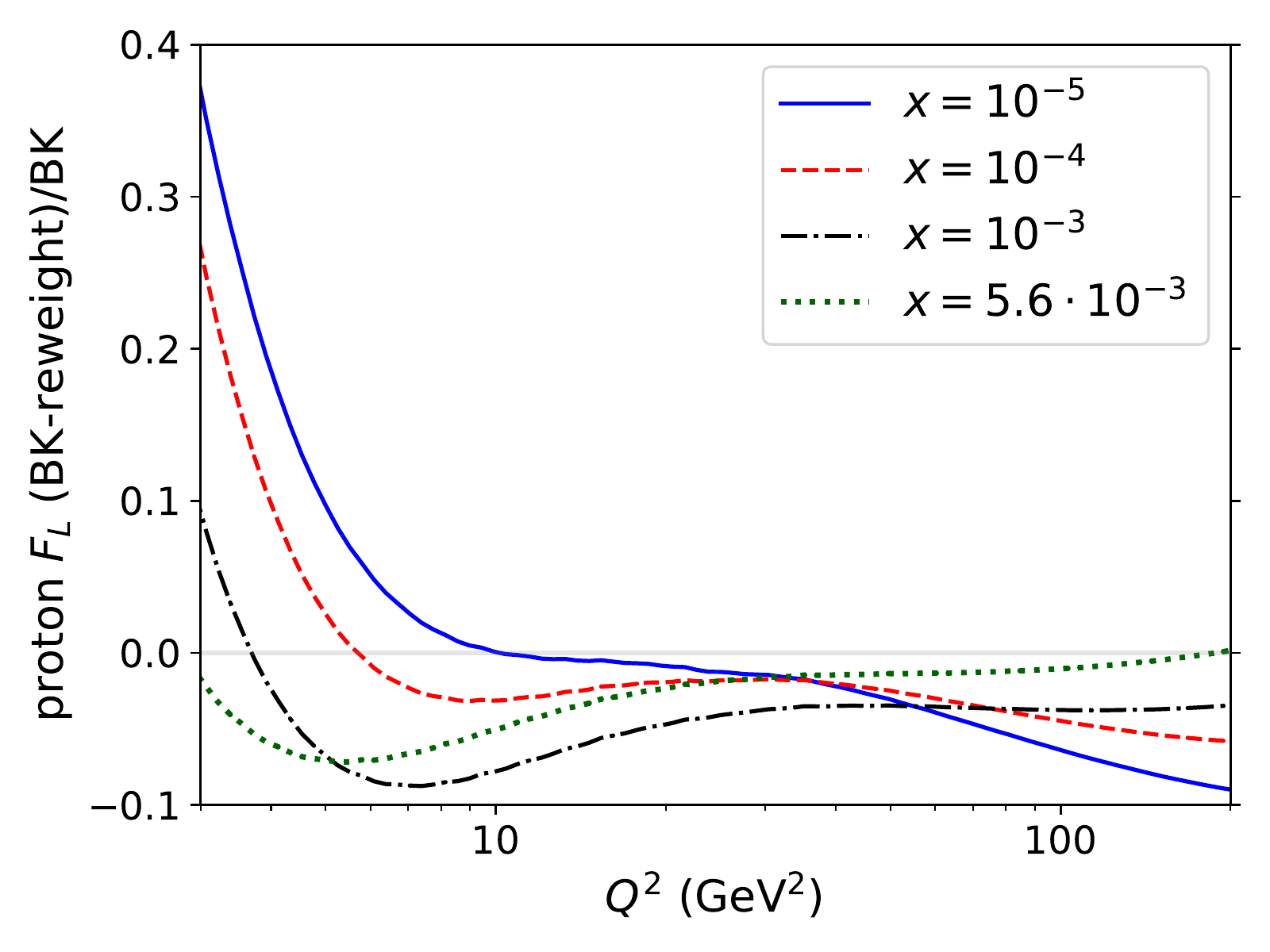}
}
\caption{The relative difference $(F_{2,{\rm L}}^{\rm BK}-F_{2,{\rm L}}^{\rm Rew})/F_{2,{\rm L}}^{\rm BK}$ between the BK predictions and the matched PDF predictions for $F_2$ (a) and $F_{\rm L}$ (b) for proton shown as a function of $Q^2$ for four different $x$ values.}
\label{fig:1DF2FLproton}
\end{figure*}

It is useful to define the so-called effective number of replicas $N_{\rm eff}$, which serves as an proxy to the number of replicas with a significant weight~\cite{Ball:2010gb, Paukkunen:2014zia}, 
\begin{equation}
\label{eq:Neff}
   N_{\rm eff} = \exp{\frac{1}{N_{\rm rep}}\sum_{k=1}^{N_{\rm rep}}\omega_k\ln(\frac{N_{\rm rep}}{\omega_k})}.
\end{equation}
In the present analysis we use $N_{\rm eff}$ to choose an appropriate value for the error parameter $\delta_{\rm BK}$: If $\delta_{\rm BK}$ is too low, $N_{\rm eff} \approx 1$ (for large  $N_{\rm rep}$) and the procedure picks up only a single replica irrespectively of how well it fits with the matching values. On the other hand, if $N_{\rm eff} \sim N_{\rm rep}$ all replicas have the equal weight and the reweighting does nothing, i.e., $\delta_{\rm BK}$ is too high. We have found that iteratively adjusting $\delta_{\rm BK}$ such that $N_{\rm eff}\approx 10$ is a good strategy for finding a set of PDFs that matches the given boundary conditions, i.e., structure functions from the BK framework.
In order to obtain $N_{\rm eff}\approx 10$ we have fixed 
$\delta_{\text{BK}}=4.5$ for the proton $F_2$, $\delta_{\text{BK}}=11.5$ for the proton $F_{\rm L}$, 
$\delta_{\text{BK}}=39.5$ for the nuclear $F_2$, and 
$\delta_{\text{BK}}=46$ for the nuclear $F_{\rm L}$. For the nuclear reweighting we use  $N_\text{data}=138$, and for the proton $N_\text{data}=125$.

\section{Results}
\label{sec:results}

\subsection{Proton}

The structure functions $F_2$ and $F_{\rm L}$ for the proton before and after the reweighting on the $Q^2=10Q_s^2(x)$ line are shown in Figs.~\ref{subfig:F2protonF2weight} and \ref{subfig:FLprotonFLweight}. The reweighting is done separately for $F_2$ and $F_{\rm L}$, as also in reality these two quantities will be measured in different kinematical domains and with a different experimental precision.  
The structure functions obtained after the reweighting can be seen to match very well to the BK results. This was to be expected since the proton PDFs and the initial condition for the BK evolution are fitted to the same precise HERA data at $x\gtrsim 10^{-4}$, and the central NNPDF3.1 results are already very close to the BK values to begin with in this domain. However, a nearly perfect agreement with the BK results is obtained also at $x\lesssim 10^{-4}$. All in all, the matching procedure  is thus found to work extremely well here.

Next we study how the differences in the BK vs. DGLAP dynamics become visible when we move away from the $Q^2\approx 10Q_s^2(x)$ line. In Figs.~\ref{subfig:HmapF2protonF2weight} and~\ref{subfig:HmapFLprotonFLweight} we show the relative difference
\begin{equation}
\label{eq:F2FL_difference}
    \frac{F_{2,L}^\text{BK} - F_{2,L}^\text{Rew}}{F_{2,L}^{\text{BK}}}
\end{equation}
as a function of both $x$ and $Q^2$, where $F_{2,L}^\text{Rew}$ refers to the corresponding structure function calculated using the reweighted PDFs. The points used in the reweighting  are also indicated in these figures. One-dimensional projections of the same quantity are plotted at fixed values of $x$ in Fig.~\ref{fig:1DF2FLproton}.

For the $F_2$ structure function shown in Fig.~\ref{subfig:HmapF2protonF2weight} the differences remain very small, at most at a few-percent level almost everywhere in the studied $x,Q^2$ range, except in the high-$x$, high $Q^2$ and low-$x$, low $Q^2$ corners. This is better visible in Fig.~\ref{subfig:1DF2protonF2weight} where we show the relative differences as a function of virtuality $Q^2$ at four different $x$ values from $x=5.6 \times 10^{-3}$ (largest $x$ for which $Q^2=10Q_s^2(x) \ge Q_0^2$, where $Q_0^2$ is the initial scale in the NNPDF3.1 PDF set) to $x=10^{-5}$. The smallest $x$ values in our plots are beyond reach for the EIC, which will collide electrons with energies $5-18$ GeV on protons and nuclei with energies 250 and 100 GeV/nucleon respectively, resulting in a kinematic reach (at $Q^2=10$ GeV$^2$) down to $x \sim 10^{-3}$~\cite{AbdulKhalek:2021gbh}. Smaller $x$ values could be probed at the LHeC (50 GeV electrons on $Z/A\times 7$ TeV/nucleon protons and nuclei) whose kinematic reach goes down to $x \sim 10^{-5}$~\cite{LHeC:2020van} and at the FCC-he~\cite{FCC:2018byv} (60 GeV electrons on $Z/A\times 50$ TeV/nucleon protons and nuclei) whose kinematic coverage extends to even lower $x$. We see that around $x \sim 10^{-4}$ the $Q^2$ dependencies are nearly equal in both frameworks. In the higher-$x$ region the BK equation predicts a stronger $Q^2$ dependence than the DGLAP equation, while in the $x\lesssim 10^{-4}$ region the BK dynamics results with a weaker $Q^2$ dependence than what the DGLAP equation predicts. As a result, at fixed $Q^2\sim 10\ \gev^2$ the relative difference changes sign as a function of $x$. Since the relative differences remain at a few-percent level, a very precise determination of the proton $F_2$ is required in order to distinguish between the two physical pictures in a statistically meaningful manner.

The differences between the BK and DGLAP dynamics are more clearly visible in the case of the structure function $F_{\rm L}$. This can be seen from Fig.~\ref{subfig:HmapFLprotonFLweight} and Fig.~\ref{subfig:1DFLprotonFLweight} which show the analogous plots for $F_{\rm L}$ that were above discussed for $F_2$. There are now larger differences even within the HERA kinematics as the $F_{\rm L}$ data from HERA are rather scarce. 
The DGLAP evolved $F_{\rm L}$ shows generically a significantly stronger $Q^2$ dependence in comparison to the BK evolved $F_{\rm L}$. The $x$ dependencies at fixed $Q^2$ are now also clearly different, particularly at $Q^2 \lesssim 10\ \gev^2$ where the BK evolution predicts a stronger $x$ dependence than what the reweighted $F_{\rm L}$ with DGLAP dynamics has. In the EIC kinematics the relative differences are maximally $\sim 10\%$. 
In the LHeC kinematics differences can be up to $\sim 40\%$ which would be easier to resolve.

Note that for the proton $F_2$ in Fig. \ref{subfig:1DF2protonF2weight} there is a small $b$ quark threshold effect in the FONLL-B scheme at $Q^2=m_b^2 \approx 24 \ \rm GeV^2$.

\begin{figure*}[ht]
\centering
\subfloat[$F_2$\label{subfig:F2AuF2weight}]{%
    \includegraphics[width=\columnwidth]{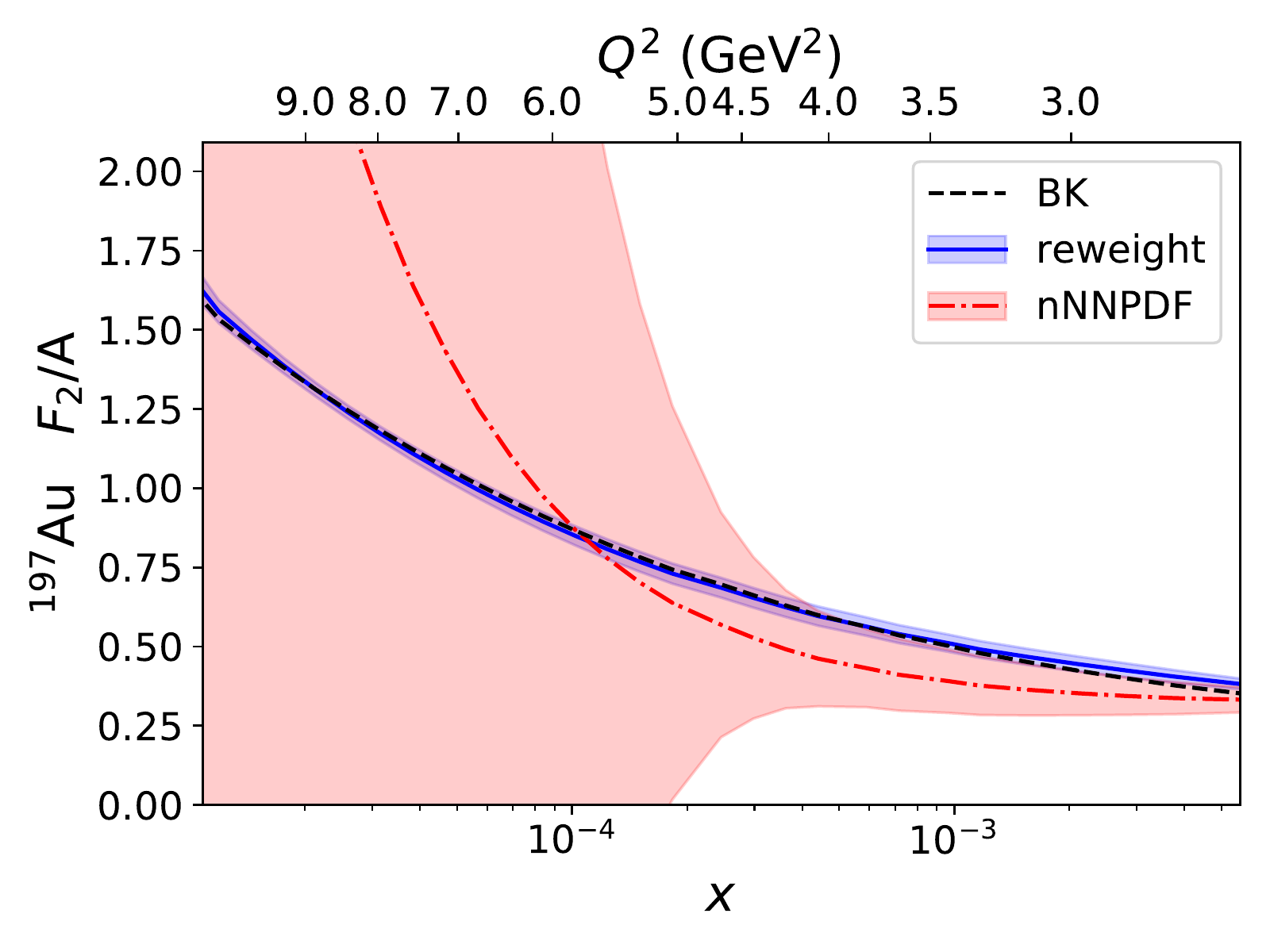}
}
\hfill
\subfloat[$F_{\rm L}$\label{subfig:FLAuFLweight}]{%
    \includegraphics[width=\columnwidth]{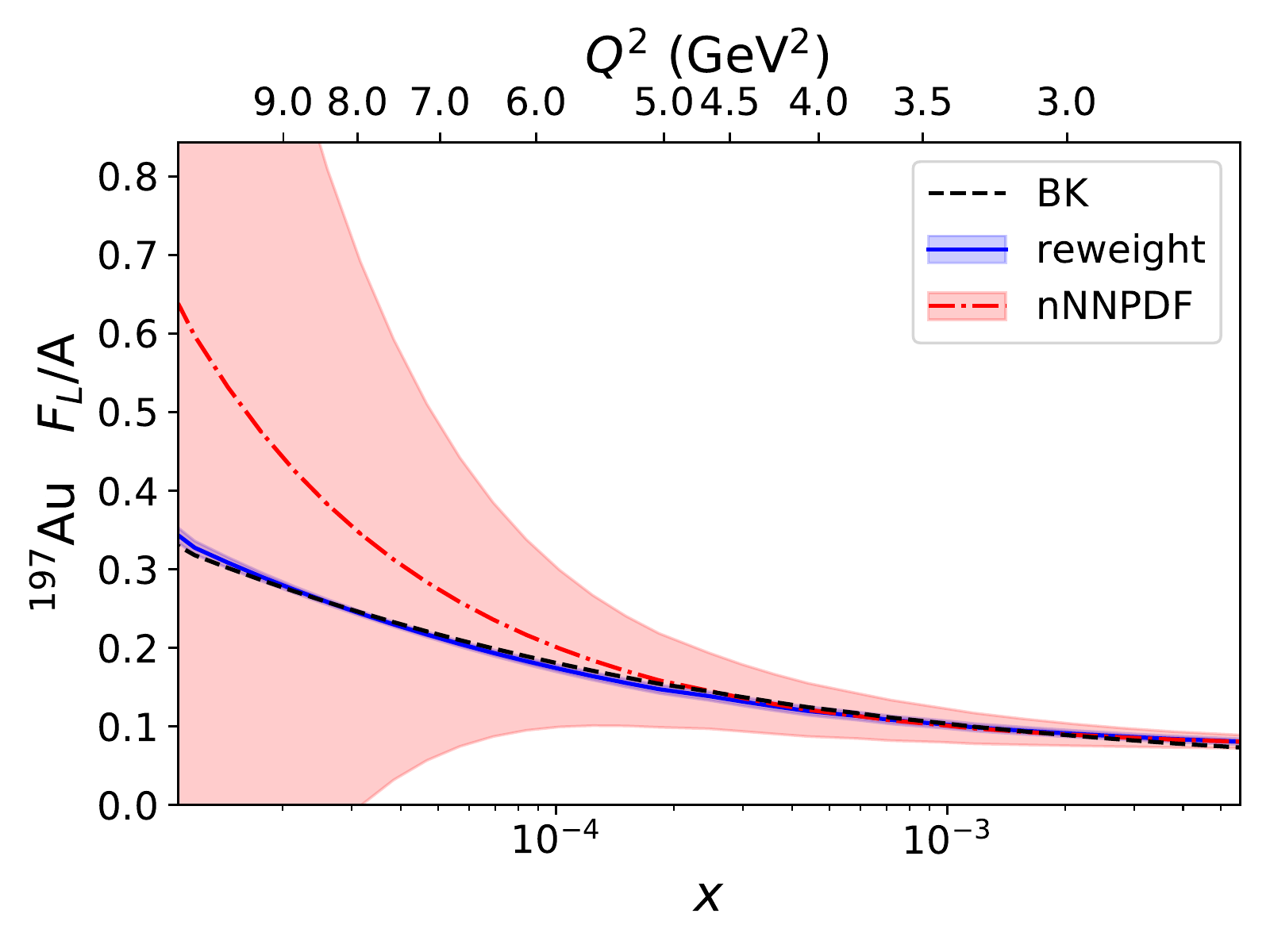}
}
\caption{
The $F_2$ (a) and $F_{\rm L}$ (b) structure functions for ${}^{197}\mathrm{Au}$ as a function of $x$ at $Q^2=10Q_s^2(x)$. The black dashed curve shows the BK predictions, the red dashed-dotted curve with the red error band the original NNPDF3.1 PDF predictions, and the blue solid curve with a light-blue errorband the PDF predictions after the matching. 
}
\label{fig:F2FLAu}
\end{figure*}

\begin{figure*}[ht]
\centering
\subfloat[$F_2$\label{subfig:HmapF2AuF2weight}]{%
\includegraphics[width=\columnwidth]{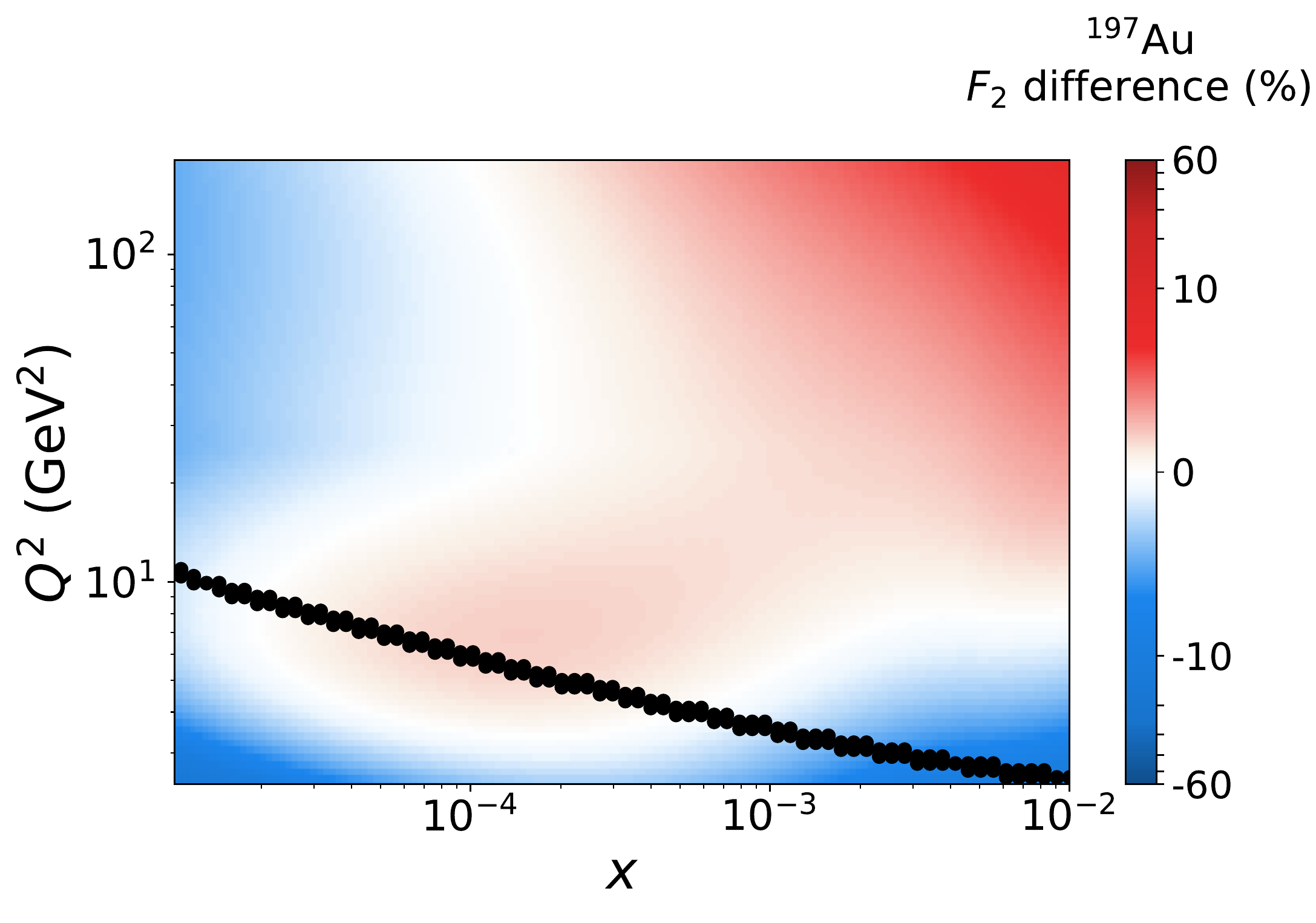}
}
\hfill
\subfloat[$F_{\rm L}$\label{subfig:HmapFLAuFLweight}]{%
    \includegraphics[width=\columnwidth]{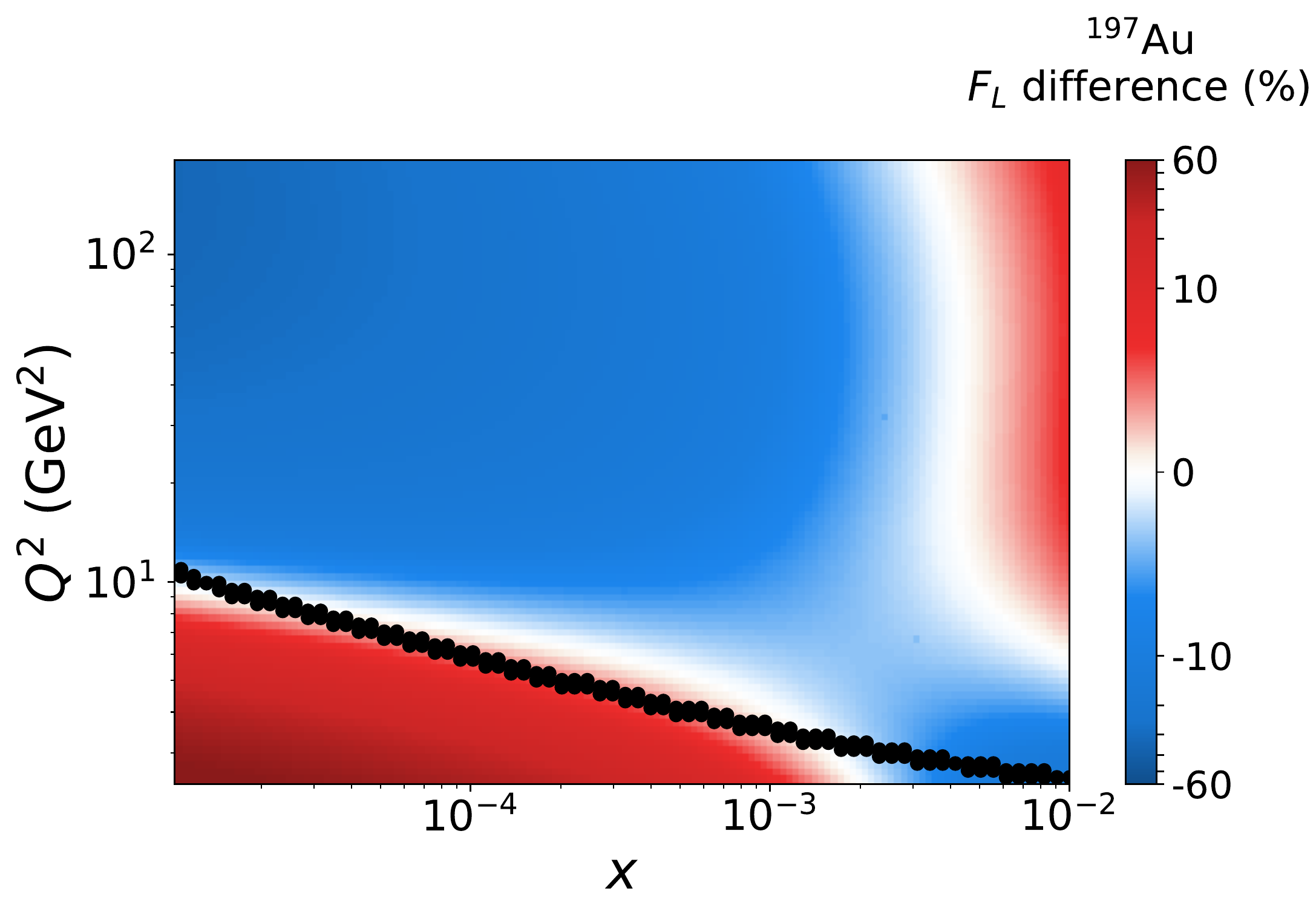}
}
\caption{
Relative difference $(F_{2,L}^{\rm BK}-F_{2,L}^{\rm Rew})/F_{2,L}^{\rm BK}$ between the BK structure functions and the matched $F_2$ (a) and $F_{\rm L}$ (b) for ${}^{197}\mathrm{Au}$ as a function of $x$ and $Q^2$. The color scale/axis goes in a linear scale from $-10\, \%$ to $10\, \%$ and in a logarithmic scale outside that range. The black dots indicate the matching points.}
\label{fig:HmapF2FLAu}
\end{figure*}

\begin{figure*}[ht]
\centering
\subfloat[$F_2$\label{subfig:1DF2AuF2weight}]{%
    \includegraphics[width=\columnwidth]{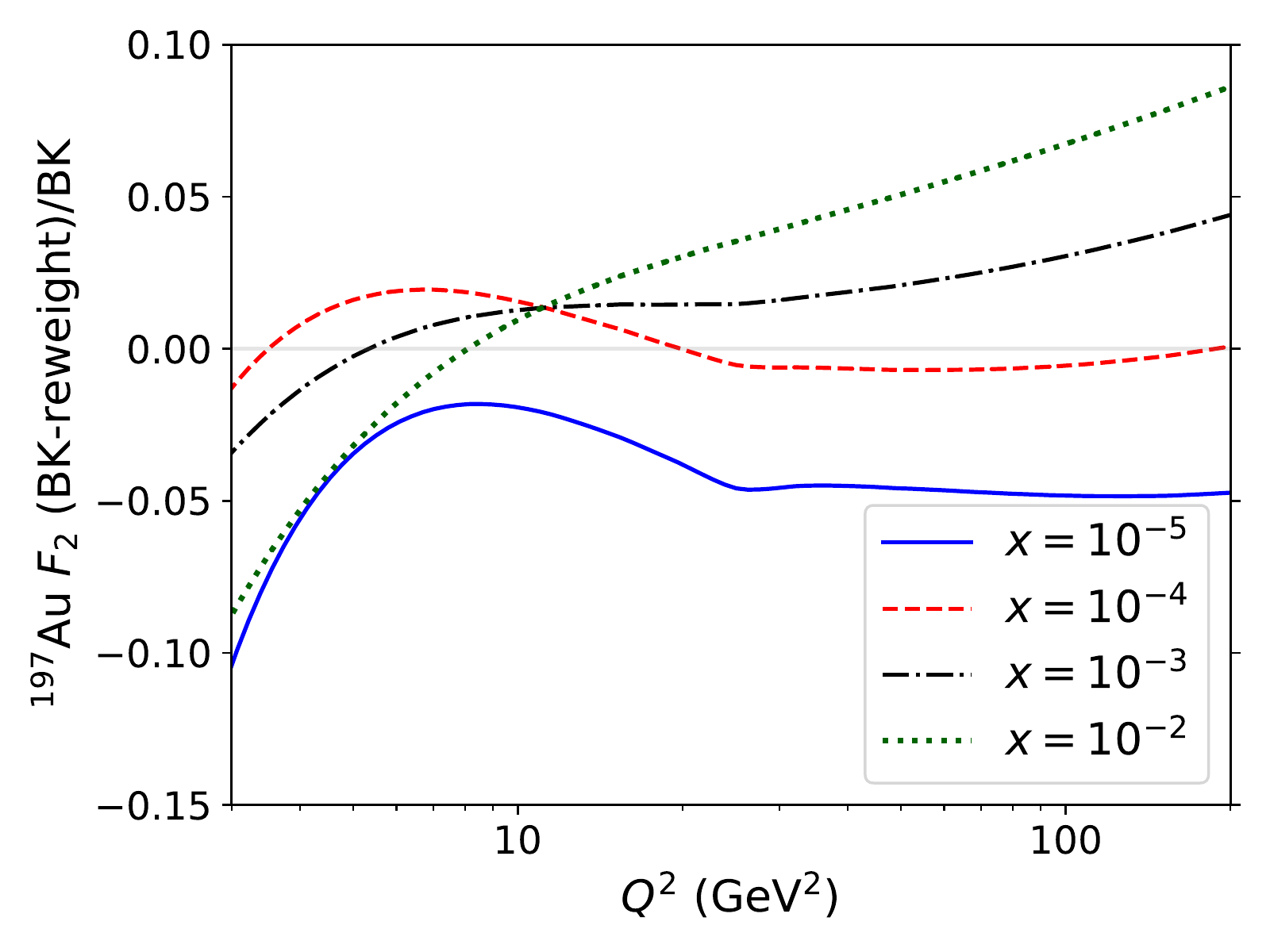}
}
\hfill
\subfloat[$F_{\rm L}$\label{subfig:1DFLAuFLweight}]{%
    \includegraphics[width=\columnwidth]{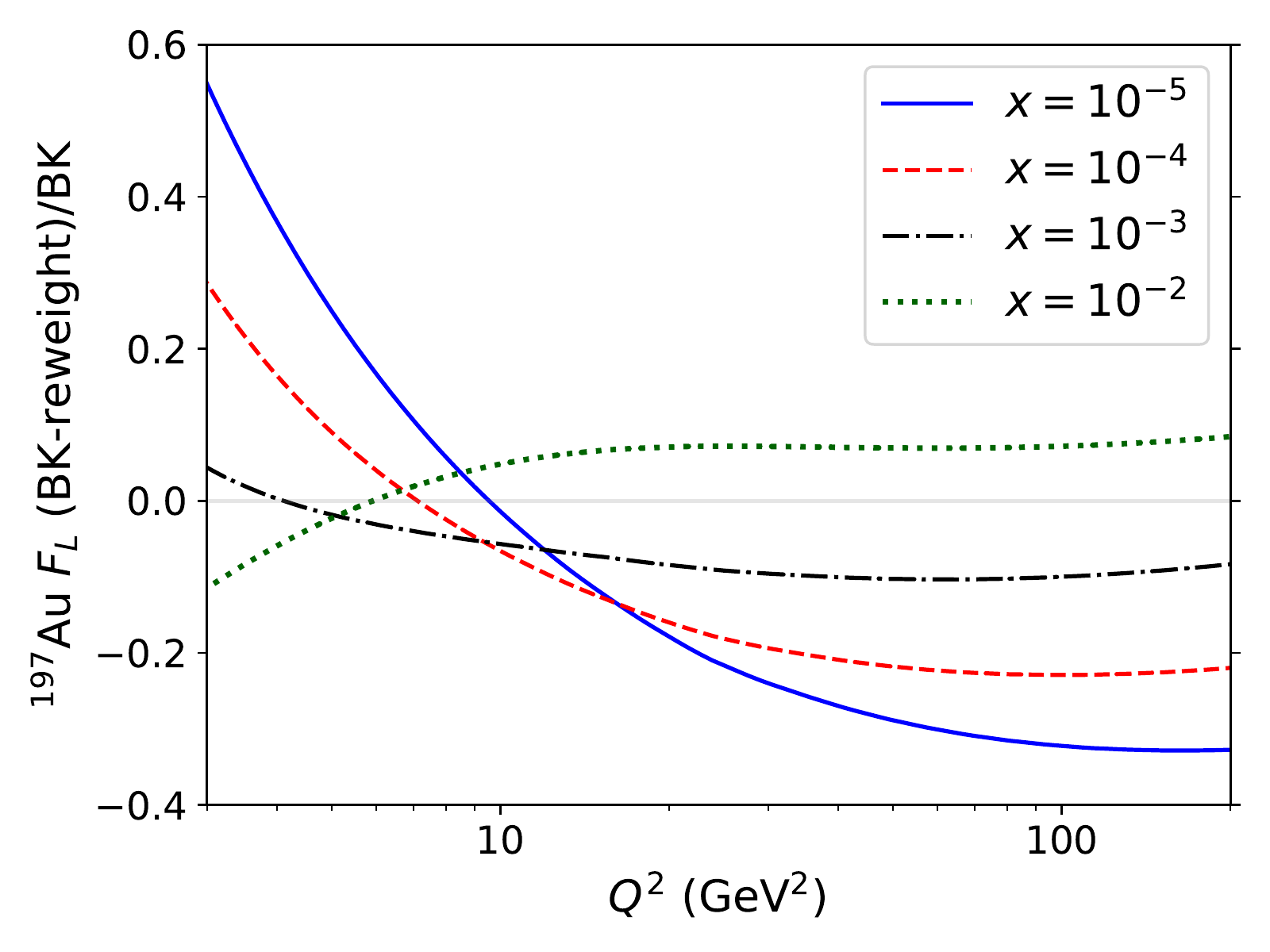}
}\caption{
The relative difference $(F_{2,{\rm L}}^{\rm BK}-F_{2,{\rm L}}^{\rm Rew})/F_{2,{\rm L}}^{\rm BK}$ between the BK predictions and the matched PDF predictions for $F_2$ (a) and $F_{\rm L}$ (b) for ${}^{197}\mathrm{Au}$ shown as a function of $Q^2$ for four different $x$ values. }
\label{fig:1DF2FLAu}
\end{figure*}

\subsection{Heavy Nucleus}

Next we study the structure functions $F_2$ and $F_{\rm L}$ for the ${}^{197}\mathrm{Au}$ nucleus. The procedure here is the same as for protons above: first we match the nuclear PDF predictions with the BK predictions in the region $10Q_s^2(x)<Q^2<11 Q_s^2(x)$, where $Q_s$ is the same proton saturation scale as what was used above. As the saturation scale of a large nucleus is generically just a few times larger than that of the proton (see e.g. Ref.~\cite{Lappi:2013zma}), at $Q^2=10Q_s^2(x)$ the non-linear effects should still be relatively small and both frameworks are expected to be applicable.

The nuclear structure functions $F_{2,{\rm L}}$ at $Q^2=10Q_s^2(x)$ before and after reweighting are shown in Figs.~\ref{subfig:F2AuF2weight} and~\ref{subfig:FLAuFLweight}. As one can see, this time the reweighting is not as successful as it was in the case of proton. The proton and nuclear PDF sets have exactly the same amount of Monte Carlo replicas but, as there are much fewer experimental constraints for the heavy-nucleus structure functions, there are larger variations between the different replicas (as illustrated by the large uncertainty band before reweighting). Consequently, the remaining uncertainty band calculated with $N_{\text{eff}}=10$ effective replicas can be much larger than with protons, and a more precise matching would require an even larger number of replicas. Thus the larger error band after reweighting is mostly just a technical limitation, not a physics effect in itself. A partial explanation for the difference between the central values before and after reweighting is that  the BK prediction is not constrained by experimental data on hard processes involving heavy nuclei, which do affect the  nuclear-PDF predictions.

Next we again study the relative differences between the DGLAP- and BK-evolved structure functions, defined by Eq.~\eqref{eq:F2FL_difference}, on the $x,Q^2$ plane. The results are shown in Fig.~\ref{subfig:HmapF2AuF2weight} for $F_2$ and in Fig.~\ref{subfig:HmapFLAuFLweight} for $F_{\rm L}$. In comparison to the proton results shown in Figs.~\ref{subfig:HmapF2protonF2weight} and~\ref{subfig:HmapFLprotonFLweight} we find much larger differences with a heavy nucleus. This is expected, as non-linear dynamics not present in the DGLAP evolution is enhanced in heavy nuclei
by a factor $\sim A^{1/3}$. 
  The fact that the results obtained with the matched PDFs do not exactly agree with the BK predictions results leads to the line of agreement between the two approaches (white in the figure) not exactly aligning with the points where the matching is done. 
However,
the differences between the two evolution equations when moving away from the $Q^2=10Q_s^2$ line are still very clear.

Again, we see more dramatic differences in $F_{\rm L}$ compared to $F_2$. For $F_2$ the cleanest systematical effect is that the BK evolution predicts slower $x$ evolution at all $Q^2$, and this difference in the evolution speed is much larger than with proton targets. In the case of $F_{\rm L}$ the DGLAP evolution results in a significantly faster $Q^2$ evolution except at the highest $x\sim 10^{-2}$, and the $x$ dependence from BK evolution is again generically slower. The fact that BK evolution generically results in a slower $x$ dependence with heavy nuclei is expected, as the effect of the non-linear contributions in the BK evolution is to slow down the evolution speed.

To quantify more accurately the expected magnitude of the deviations, we show in Figs.~\ref{subfig:1DF2AuF2weight} and \ref{subfig:1DFLAuFLweight} the relative difference from Figs.~\ref{subfig:HmapF2AuF2weight} and~\ref{subfig:HmapFLAuFLweight} for fixed $x$ values in the EIC and LHeC/FCC-he kinematics as a function of $Q^2$. The reweighted $F_2$ values differ from the BK-evolved results less than $10\,\%$ in the studied kinematical domain, which implies that also with heavy nuclei the $F_2$ should be measured at a few-percent precision in order to be sensitive to the differences in the evolution. For $F_{\rm L}$ the differences are again significantly larger and up to $60\%$ in the LHeC/FCC-he kinematics and up to $15\%$ in the EIC kinematics.
Similarly to protons, a bottom quark mass threshold effect in the FONLL-B scheme is seen in the nuclear $F_2$ in Fig.~\ref{subfig:1DF2AuF2weight} at $Q^2\approx 24\ \rm GeV^2$.

\subsection{Effects on PDFs}

\begin{figure*}[ht]
\centering
\subfloat[gluon\label{fig:gluon_protonreweight}]{%
\includegraphics[width=\columnwidth]{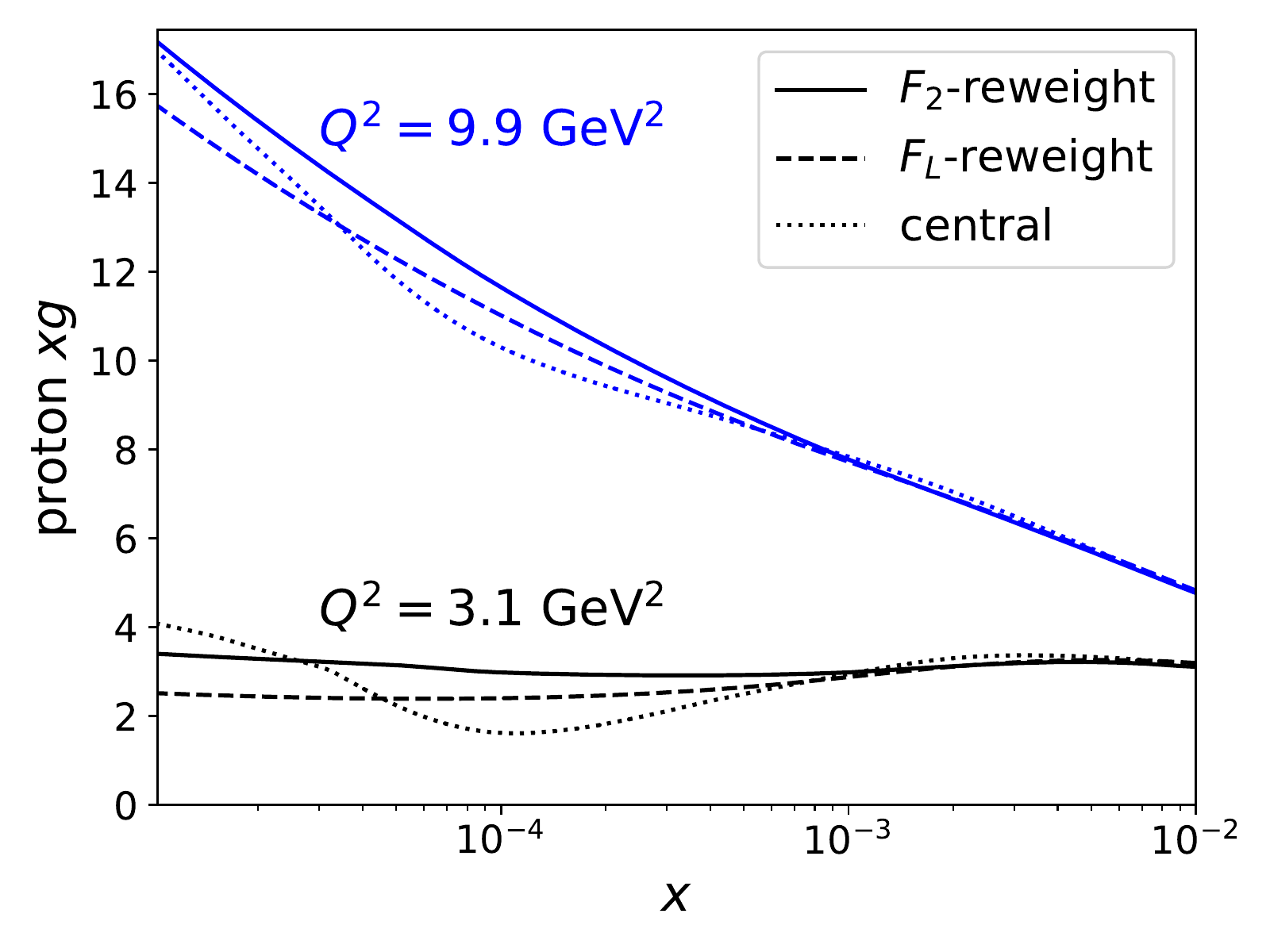}
}
\subfloat[up quark\label{fig:uquark_protonreweight}] {%
\includegraphics[width=\columnwidth]{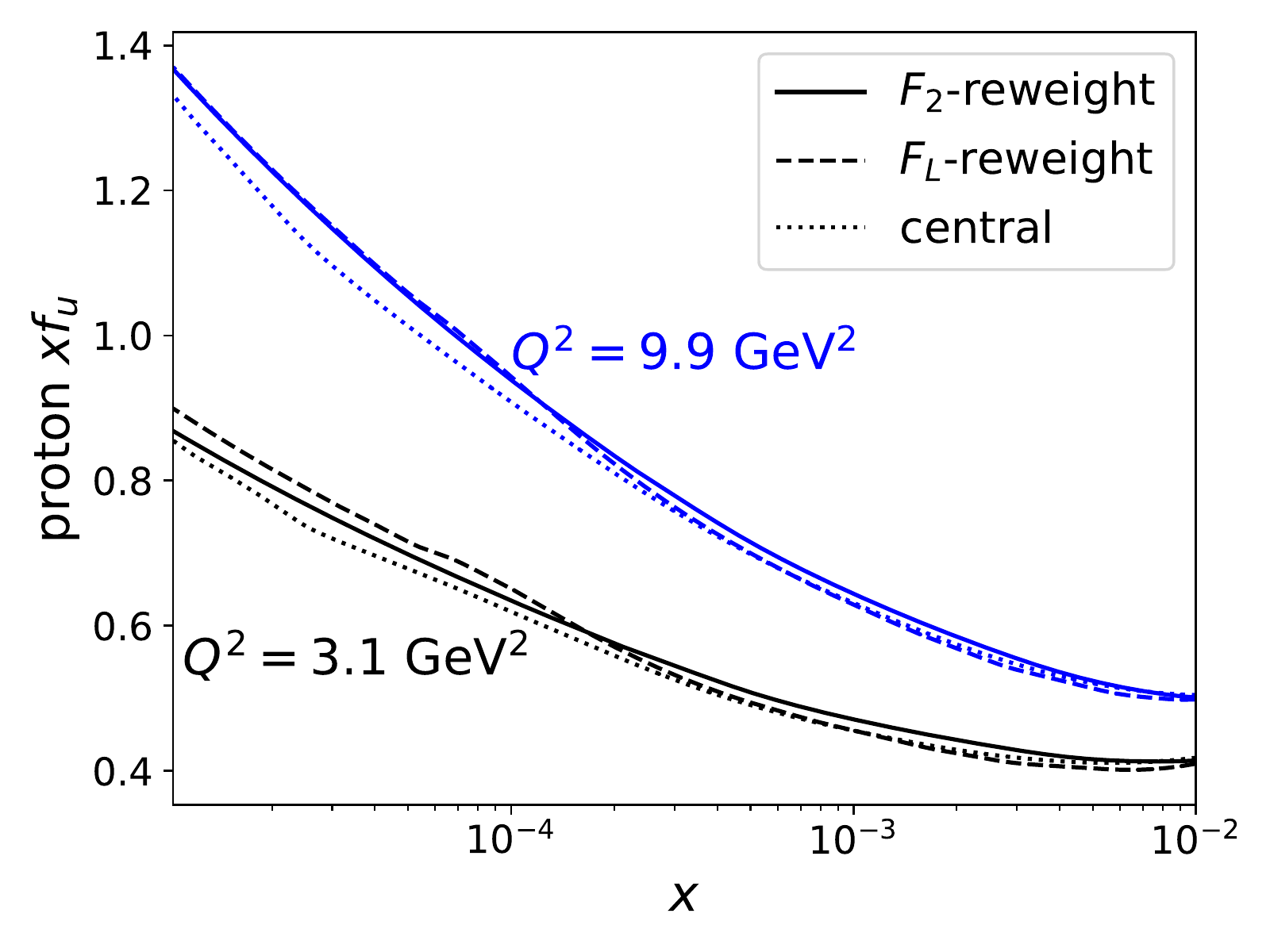}
}
\caption{Reweighted (solid and dashed curves) and central (dotted curves) gluon (a) and up-quark (b) NNPDF3.1 proton PDFs as a function of $x$ at low (black) and moderate $Q^2$ (blue). Results matched with the BK predictions for $F_2$ (solid curves) and $F_{\rm L}$ (dashed curves) are shown separately.
}
\end{figure*}

\begin{figure*}[ht]
\centering
\subfloat[gluon\label{fig:gluon_Aureweight}]{%
\includegraphics[width=\columnwidth]{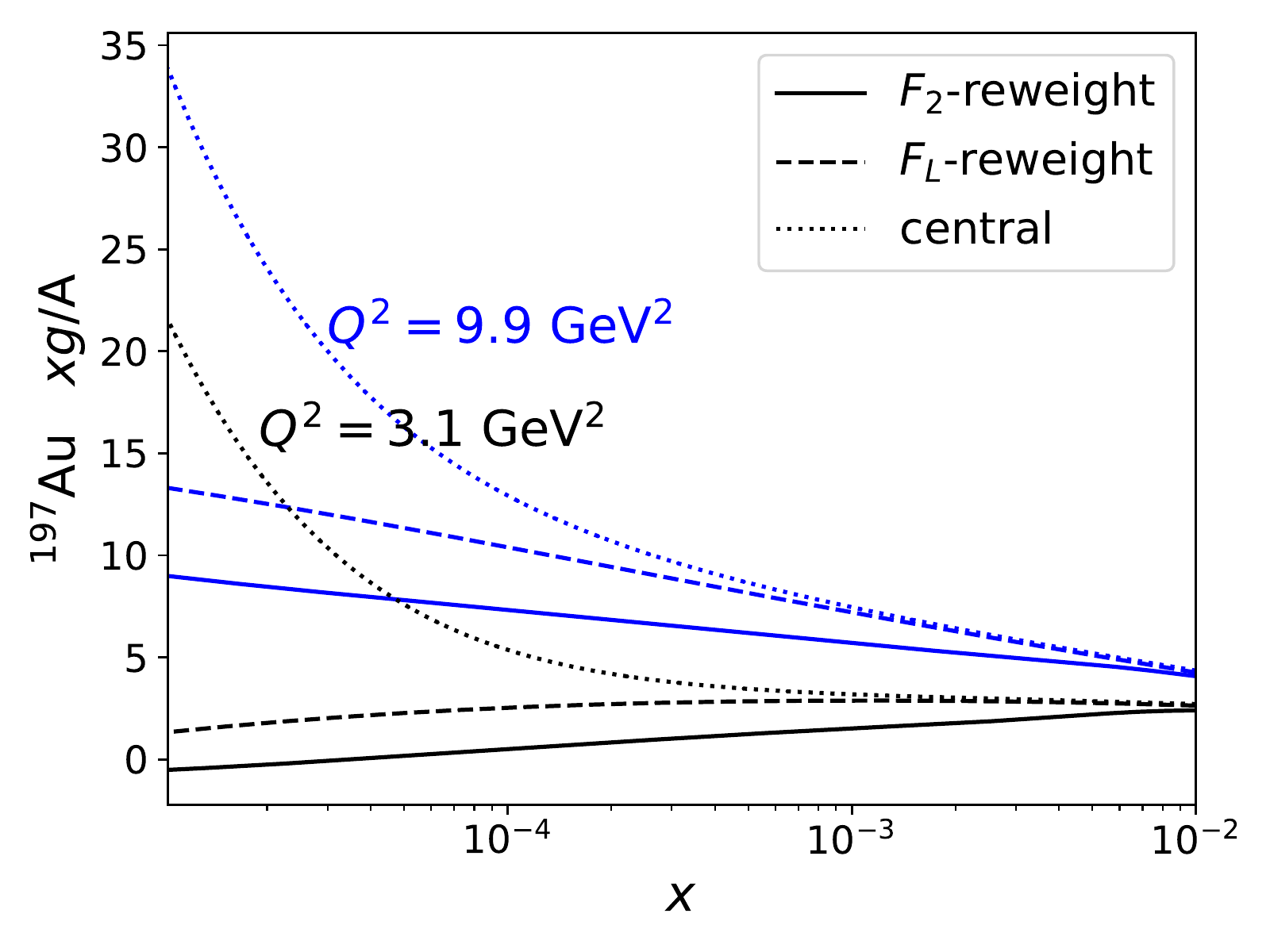}
}
\subfloat[up quark\label{fig:uquark_Aureweight}]{%
\includegraphics[width=\columnwidth]{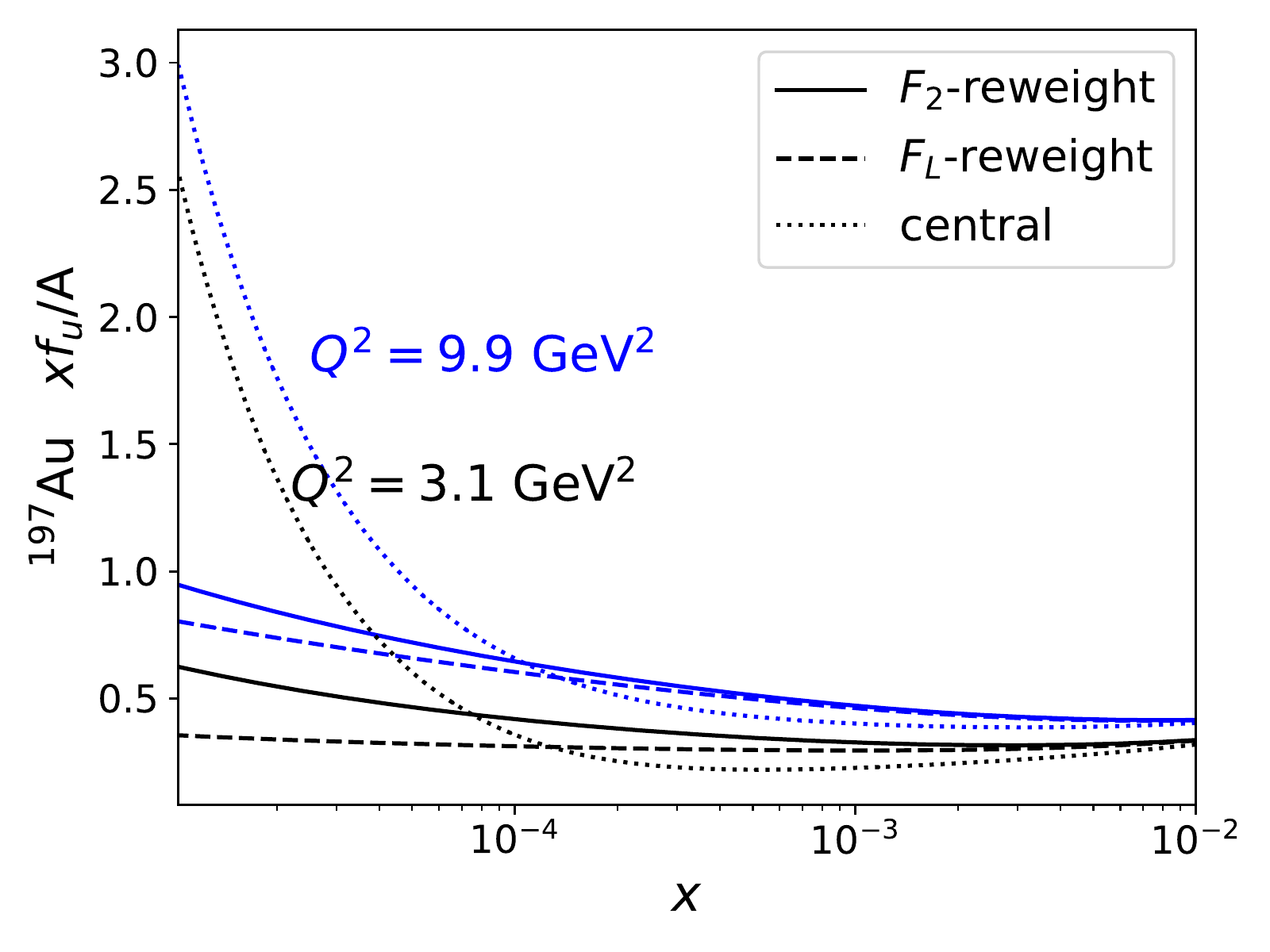}
}
\caption{Reweighted (solid and dashed curves) and central (dotted curves) gluon (a) and up-quark (b) nNNPDF2.0 ${}^{197}\mathrm{Au}$ PDFs as a function of $x$ at low (black) and moderate $Q^2$ (blue). Results matched with the BK predictions for $F_2$ (solid curves) and $F_{\rm L}$ (dashed curves) are shown separately. 
}
\end{figure*}

The matching process results in a new PDF set that is constrained by the BK evolved structure functions on the $Q^2=10\dots 11 Q_s^2(x)$ line. In this section we study how these constraints originating from the non-linear dynamics included in the BK evolution alter the parton distribution functions. The coefficients $\omega_k$ defined in Eq.~\eqref{eq:omega} determined in the reweighting process for the applied PDF sets are made available in the Supplementary material. Using these coefficients and the publicly available NNPDF3.1 or nNNPDF2.0 sets it is then possible to construct DGLAP evolved parton distribution functions that match the saturation model predictions at moderate $Q^2$ and are compatible with the neutral-current HERA data. 

First we illustrate here the effect of reweighting on the proton parton distribution functions. The results are shown in Figs.~\ref{fig:gluon_protonreweight} and \ref{fig:uquark_protonreweight} for the gluon and up-quark distributions at two different virtualities. In order to study how complementary the observables $F_2$ and $F_{\rm L}$ are, we separately show the results with matching to $F_2$ and matching to $F_{\rm L}$. The central set values from the NPDF3.1 are also shown for comparison. Comparing Fig.~\ref{fig:gluon_protonreweight} to \ref{fig:uquark_protonreweight} we find the reweighting to have a slightly stronger effect on the  gluon distribution compared to up quarks, but in general the reweighting has only a modest effect on the proton PDFs. This is expected, as the PDFs are fitted to the same HERA data that is used to constrain the BK boundary conditions.  Whether $F_2$ or $F_{\rm L}$ is used in reweighting has only a small effect on the determined reweighted PDFs. Thus, we do not expect to see strong tensions when measurements from the EIC or LHeC/FCC-he are eventually used to disentangle  the BK and DGLAP dynamics.

The reweighted nuclear up-quark and gluon distributions are shown in Figs.~\ref{fig:gluon_Aureweight} and \ref{fig:uquark_Aureweight}. Comparing to the proton results shown in Figs.~\ref{fig:gluon_protonreweight} and \ref{fig:uquark_protonreweight} we see that nuclear PDFs are affected much more by the reweighting already in the $x\lesssim 10^{-3}$ region, which is expected, as in nNNPDF2.0 there are only few data constraints in this region. The reweighted nuclear PDFs are suppressed by a large factor compared to the central values from the nNNPDF2.0 set. Again both $F_2$ and $F_{\rm L}$ pseudodata have similar effects and as such no strong tensions with already existing data included in the nuclear PDF fits are expected in global analyses. In Fig.~\ref{fig:gluon_Aureweight} the nuclear gluon distribution, reweighted with $F_2$ data, becomes negative at small $x\lesssim 2 \cdot 10^{-5}$ and at $Q^2=3.1\ \rm GeV^2$. However, the gluon distribution is not an observable, and structure functions remain positive.

\section{Conclusions}
\label{sec:conclusions}

Understanding the importance of nonlinear effects is a major motivation for future high-energy scattering experiments. Experimental measurements have, however, a limited precision and kinematical range. It is therefore essential to quantify in a clear way the  magnitude of the effects that one is looking for. In this paper we have done this by comparing the DIS structure functions
$F_2$  and $F_{\rm L}$ calculated using the running coupling BK equation, which includes saturation effects, to collinear factorization at NLO. The approaches require a set of nonperturbative initial conditions in some initial line in the $(x,Q^2)$-plane and, given these initial conditions, make predictions in other regions. In order to compare the two approaches to evolution it is essential to differentiate between the effects of these initial conditions, and the actual dynamical predictions of the evolution. Our focus here is on the evolution dynamics. Thus we perform this comparison by choosing the nonperturbative initial conditions so that they match in a  regime where both frameworks should be valid, at virtualities somewhat above the saturation scale. Technically this is achieved by  profiting from large samples of DGLAP-evolved Monte-Carlo replicas of PDFs available from the NNPDF collaboration, and using reweighting to match them to the structure functions from BK evolution. There is already a significant agreement due to the fact that both calculations have been (for protons) fit to the same HERA data. The deviations of the two calculations, when moving away from the matching line, provide a clear assessment of the quantitative difference in the dynamics between the two.

As a result of this analysis, we see that $F_2$ has a quite weak sensitivity to the difference between the evolution equations, at least in the region where $Q^2$ is large enough for the PDF sets to be available. The relative differences in $F_2$ are at a percent level for protons, and somewhat higher for nuclei. The longitudinal structure function $F_{\rm L}$ is significantly more sensitive, but unfortunately also harder to measure as precisely. Overall the difference, as expected, is strongest at the lowest $Q^2$ values. Pushing collinear factorization to lower scales would clearly make the difference more manifest. In finer details, the behavior of the difference in the $x,Q^2$-plane also has more unexpected aspects.

We have here focused on providing a comparison of the predictions on a purely theoretical level, independently of any assumptions about a specific experiment. To turn this calculation into a more controlled estimate relevant to a specific measurement requires, in addition, a more detailed inclusion of estimated experimental errors, their correlations, the expected kinematical coverage and so on. This we leave to future work. Another interesting avenue for the future would be use this as a starting point for using the two approaches together. It should be possible to use  BK evolution to predict the behavior at relatively low $Q^2$ values and very  low $x$ and  combine this with DGLAP evolution towards higher $Q^2$. Such a combined framework would significantly extend the predictive power of weak coupling QCD by replacing a part of what are now initial conditions fitted to data by predictions from a first-principles calculation. It would be interesting to perform the matching required in such a calculation  analytically. This is, however, beyond the scope of this work. Instead we take a first step in this direction by providing, as a supplementary material to this paper, the DGLAP-evolved PDF sets for protons and gold nuclei that do match the BK-evolved cross section at our chosen matching scale.

\section*{Acknowledgments}

This work was supported by the Academy of Finland, the Centre of Excellence in Quark Matter  (project 346324), and projects  321840, 338263, 346567, and 308301.
This work was also supported by the European Union’s Horizon 2020 research and innovation programme under grant agreement No. 824093 and  the European Research Council under project ERC-2018-ADG-835105 YoctoLHC. NA acknowledges financial support by Xunta de Galicia (Centro singular de investigaci\'on de Galicia accreditation 2019-2022); the ``Mar\'\i{}a de Maeztu'' Units of Excellence program
MDM2016-0692 and the Spanish Research State Agency under project FPA2017-83814-P; European Union ERDF; and MSCA RISE 823947 ``Heavy ion collisions: collectivity and precision in saturation
physics'' (HIEIC); and the Spanish Research State Agency (Agencia
Estatal de Investigación).  The content of this article does not reflect the official opinion of the European Union and responsibility for the information and views expressed therein lies entirely with the authors.

\appendix

\section{Matching at a higher scale}
\label{app:highQmatch}
\begin{figure*}[ht] 
\centering
\subfloat[$F_2$\label{fig:F2Au_highscale}]{%
\includegraphics[width=\columnwidth]{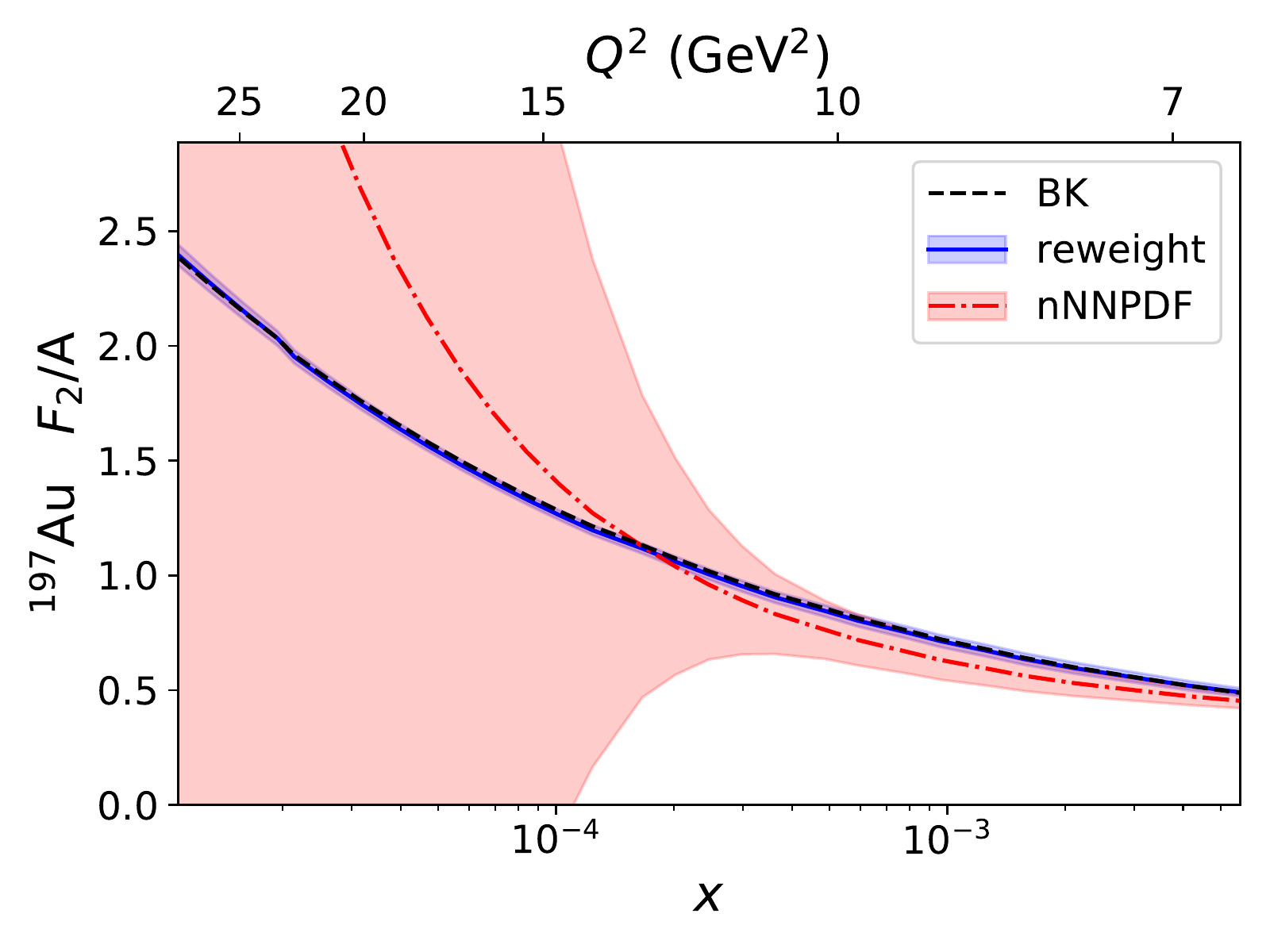}
}
\subfloat[$F_{\rm L}$\label{fig:FLAu_highscale}] {%
\includegraphics[width=\columnwidth]{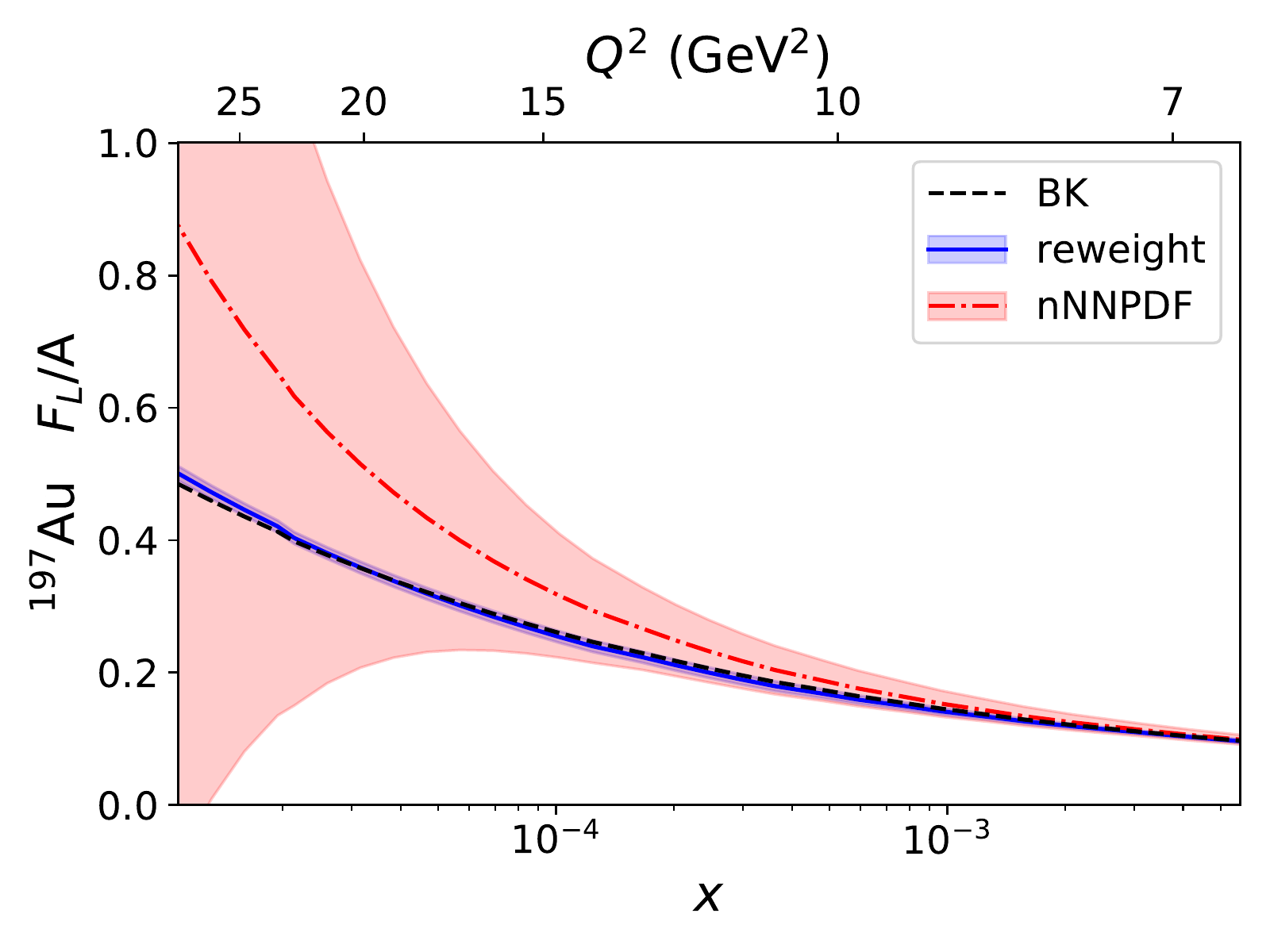}
}
\caption{
The $F_2$ (a) and $F_{\rm L}$ (b) structure functions for ${}^{197}\mathrm{Au}$ as a function of $x$ at $Q^2=25...28 Q_s^2(x)$. The black dashed curve shows the BK predictions, the red dashed-dotted curve with the red error band the original NNPDF3.1 PDF predictions, and the blue solid curve with a light-blue errorband the PDF predictions after the matching. 
}\label{fig:reweighthighQ}
\end{figure*}

\begin{figure*}[ht]
\centering
\subfloat[$F_2$\label{fig:HmapF2Au_highscale}]{%
\includegraphics[width=\columnwidth]{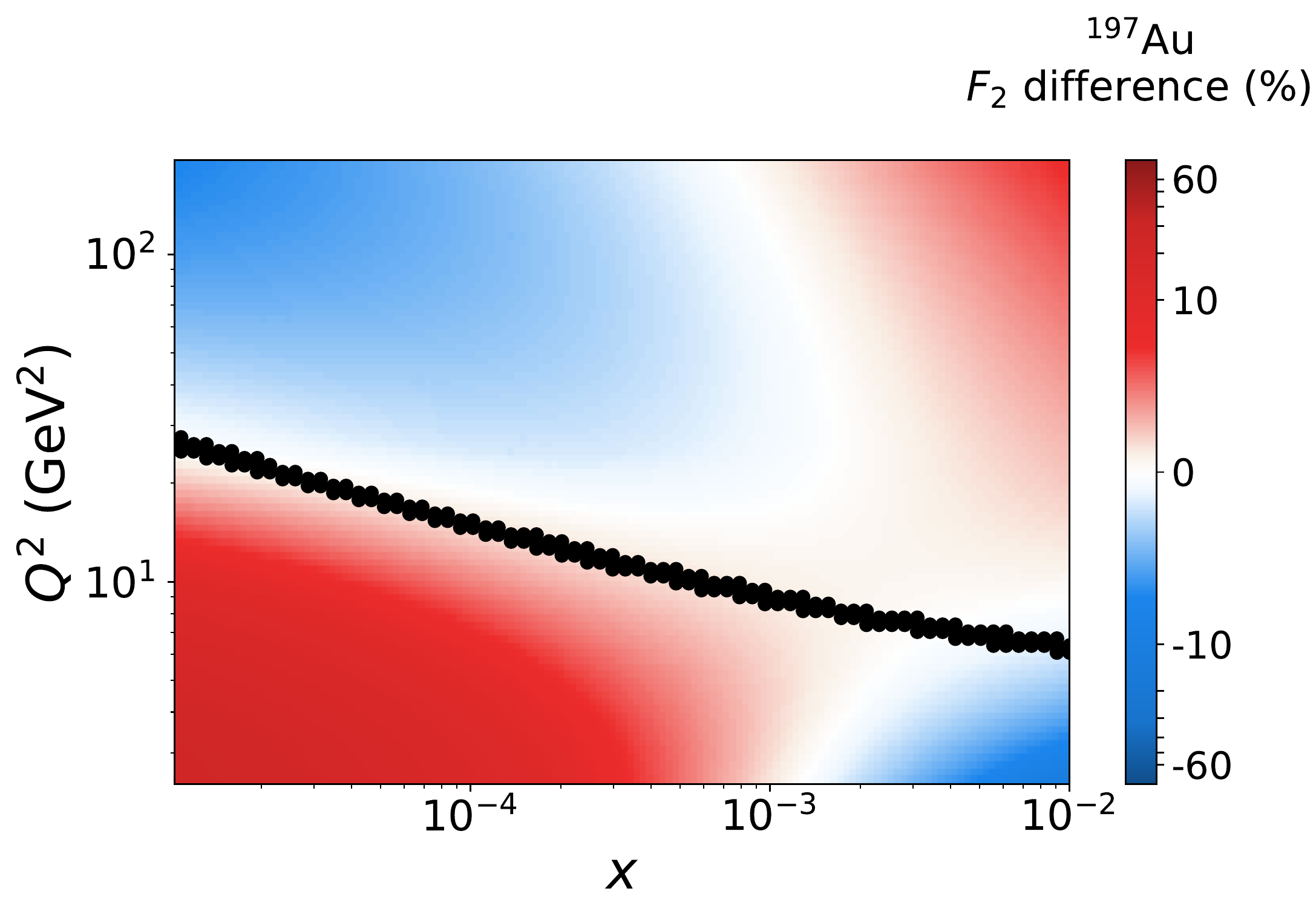}
}
\subfloat[$F_{\rm L}$\label{fig:HmapFLAu_highscale}] {%
\includegraphics[width=\columnwidth]{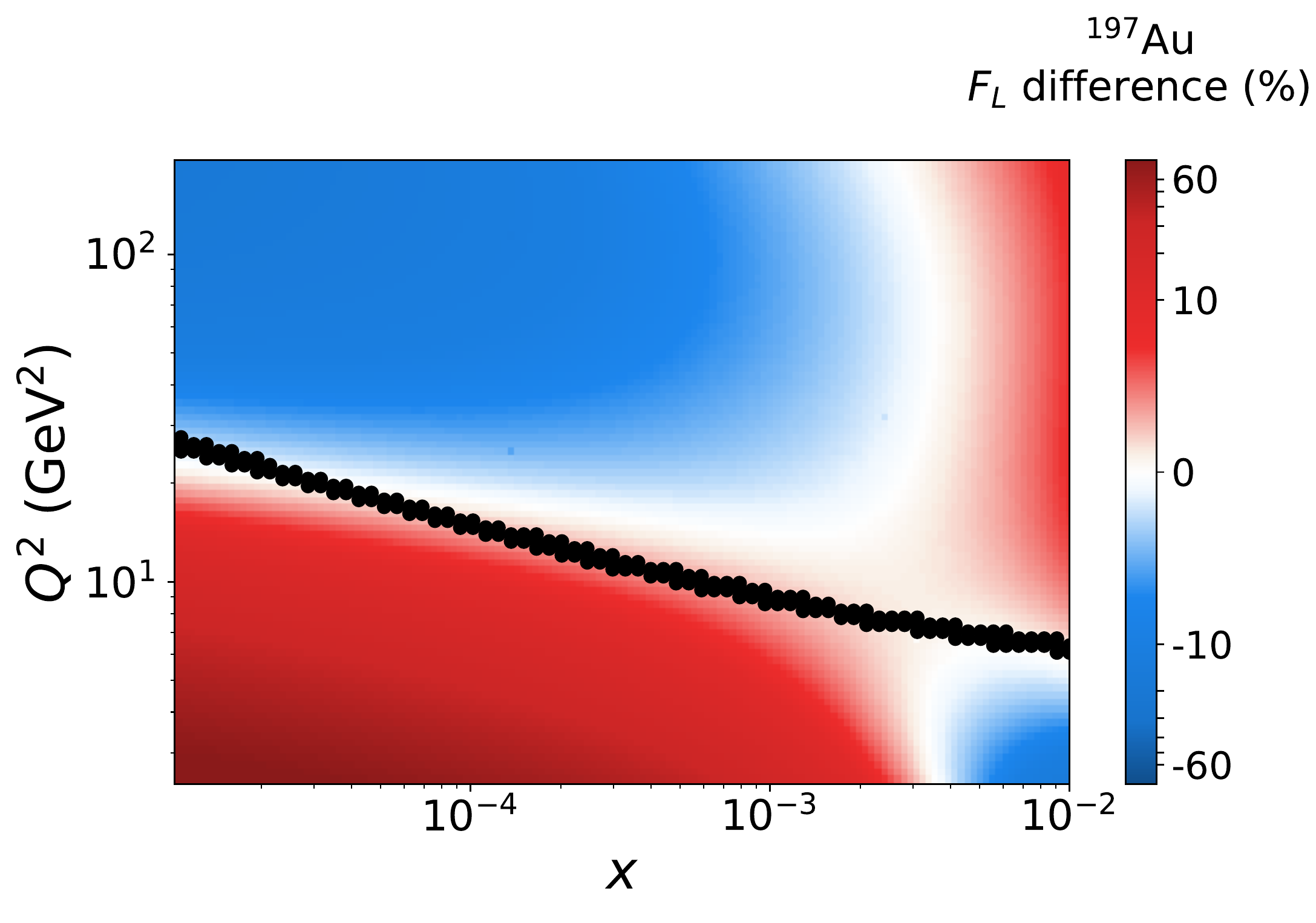}
}
\caption{
Relative difference $(F_{2,L}^{\rm BK}-F_{2,L}^{\rm Rew})/F_{2,L}^{\rm BK}$ between the BK structure functions and the matched $F_2$ (a) and $F_{\rm L}$ (b) for ${}^{197}\mathrm{Au}$ as a function of $x$ and $Q^2$. The color scale/axis goes in a linear scale from $-10\, \%$ to $10\, \%$ and in a logarithmic scale outside that range. The black dots indicate the matching points.
}\label{fig:heatmaphighQ}
\end{figure*}

\begin{figure*}[ht]
\centering
\subfloat[$F_2$\label{fig:1DF2Au_highscale}]{%
\includegraphics[width=\columnwidth]{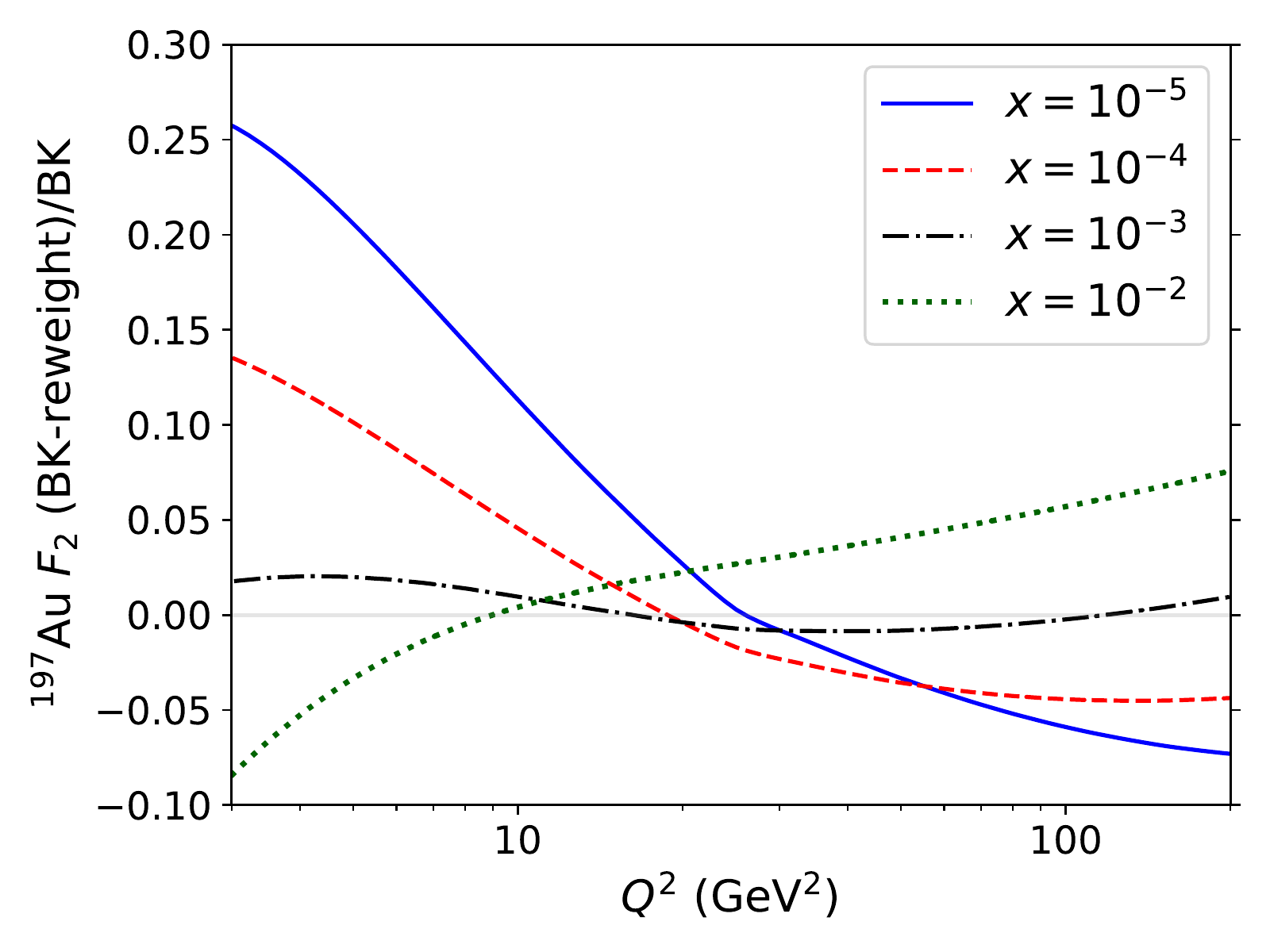}
}
\subfloat[$F_{\rm L}$\label{fig:1DFLau_highscale}] {%
\includegraphics[width=\columnwidth]{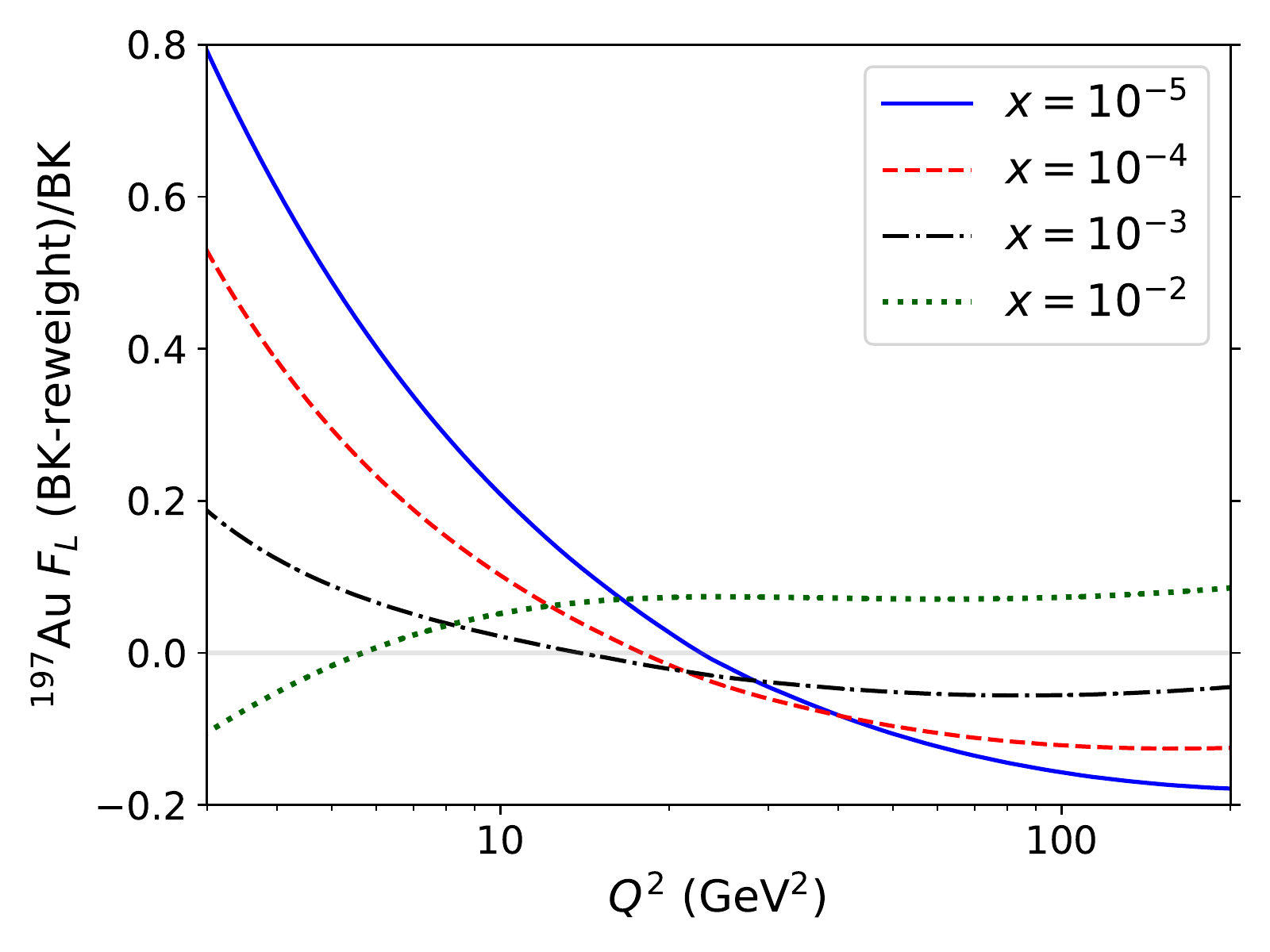}
}
\caption{
The relative difference $(F_{2,{\rm L}}^{\rm BK}-F_{2,{\rm L}}^{\rm Rew})/F_{2,{\rm L}}^{\rm BK}$ between the BK predictions and the matched PDF predictions for $F_2$ (a) and $F_{\rm L}$ (b) for ${}^{197}\mathrm{Au}$ shown as a function of $Q^2$ for four different $x$ values. 
}\label{fig:slicehighQ}
\end{figure*}

Figures \ref{fig:reweighthighQ}, \ref{fig:heatmaphighQ} and~\ref{fig:slicehighQ} show the result of applying our matching procedure  in the kinematical range $Q^2=25\dots 28 Q_s^2(x)$ for the gold nucleus. These figures should be contrasted with Figs.~\ref{fig:F2FLAu}, \ref{fig:HmapF2FLAu} and~\ref{fig:1DF2FLAu}, where the same quantities are plotted with our default matching scale. One can see that an equally good matching at this higher scale can be achieved. In this case there is much more phase space available below the matching scale, and at the lower $Q^2$ values the disagreement between the two calculations becomes quite substantial. This indicates that a factor $\sim 2.5$ in $Q^2$ is already sufficient to reveal a substantial difference between what is in some sense a backward DGLAP evolution towards lower scales, and the BK dynamics. This is, however, not relevant for an actual determination based on experimental data. First,  experimentally there is more phase space and thus more data available at the lowest scales. Second, and more important, allowing both calculations to adjust nonperturbative initial conditions to the same data will tend to match them at a scale which makes them differ less, which in this case is at smaller values of $Q^2$.

\bibliographystyle{JHEP-2modlong.bst}
\bibliography{refs}

\end{document}